\documentclass[superscriptaddress,aps,prl,reprint,twocolumn,amsmath,amssymb]{revtex4-2}
\usepackage{graphicx}
\usepackage{hyperref}
\usepackage{amssymb}
\usepackage{slashed}
\usepackage{dcolumn}
\usepackage{amsmath}
\usepackage{bm}
\usepackage{colordvi}
\usepackage{algorithm}
\usepackage{algpseudocode}
\usepackage{multirow}
\usepackage{titlesec}
\usepackage{newtxtext,newtxmath}
\usepackage{dsfont}
\usepackage{xcolor}
\usepackage{mathbbol}

\usepackage{hyperref}
\hypersetup{
    colorlinks=true,
    linkcolor=blue,
    citecolor=blue,     
    urlcolor=blue,
}

\allowdisplaybreaks

\usepackage{mathrsfs}
\makeatletter

\newcommand{\Rmnum}[1]{\expandafter\@slowromancap\romannumeral #1@}
\makeatother

\begin{document}

\title{Microscopic Phase-Transition Framework for Gate-Tunable Superconductivity in Monolayer WTe$_2$}

\author{F. Yang}
\email{fzy5099@psu.edu}

\affiliation{Department of Materials Science and Engineering and Materials Research Institute, The Pennsylvania State University, University Park, PA 16802, USA}

\author{G. D. Zhao}
\affiliation{Department of Materials Science and Engineering and Materials Research Institute, The Pennsylvania State University, University Park, PA 16802, USA}

\author{Y. Shi}
\affiliation{Department of Materials Science and Engineering and Materials Research Institute, The Pennsylvania State University, University Park, PA 16802, USA}

\author{L. Q. Chen}
\email{lqc3@psu.edu}

\affiliation{Department of Materials Science and Engineering and Materials Research Institute, The Pennsylvania State University, University Park, PA 16802, USA}

\date{\today}

\begin{abstract}
The recently reported gate-tunable superconductivity in monolayer WTe$_2$ {[Science {\bf 362}, 922 (2018); Science {\bf 362}, 926 (2018); Nat. Phys. {\bf 20}, 269 (2024); PRR {\bf 7}, 013224 (2025)]} exhibits several striking anomalies beyond the standard paradigm, including a contrasting carrier‐density dependence of the transition temperature $T_c$ in weakly and strongly disordered regimes and more surprisingly, the sudden disappearance of superconducting fluctuations below a critical carrier density. To understand these features, we go beyond  mean‐field theory and develop a microscopic framework that treats  the gap and superfluid density by explicitly and self-consistently incorporating both Nambu–Goldstone phase fluctuations and Berezinskii–Kosterlitz–Thouless fluctuations. We show that these fluctuations are minimal in the weak-disorder regime but become crucial under strong disorder, where the zero-temperature gap renormalized by NG quantum fluctuations becomes density-dependent while the BKT fluctuations drive the $T_c$ below the gap-closing temperature. Simulations within this unified framework combining with the density-functional-theory input to account for the excitonic instability quantitatively  reproduced nearly all  key experimental observations, providing a consistent understanding of reported anomalies. 

\end{abstract}

\pacs{74.20.-z,74.40.+k,74.62.-c}

\maketitle  

{\sl Introduction.---}Over the past few decades, advances in material synthesis and growth techniques have led to the discovery of numerous two-dimensional quantum materials. A particularly striking development is the observation of superconductivity~\cite{saito2016highly,qiu2021recent}, not only in conventional superconducting (SC) metals~\cite{li2023proximity,zhang2010superconductivity,falson2020type,liao2018superconductivity,zhang2015detection,briggs2020atomically} and high‐$T_c$ materials~\cite{yu2019high,he2013phase,lee2014interfacial,ge2015superconductivity,ding2016high} realized as ultrathin films and atomic sheets  but also in strictly monolayer (ML) transition‐metal dichalcogenides such as MoS$_2$~\cite{ye2012superconducting,lu2015evidence,saito2016superconductivity,costanzo2016gate}, WTe$_2$~\cite{fatemi2018electrically,sajadi2018gate,song2024unconventional,PhysRevResearch.7.013224} and NbSe$_2$~\cite{xu2015experimental,sun2016majorana,wang2017high,tsen2016nature}.  These unique ML systems often exhibit unconventional and intriguing phenomena beyond the standard paradigm. A notable example is the recently reported ML  WTe$_2$~\cite{fatemi2018electrically,sajadi2018gate,song2024unconventional,PhysRevResearch.7.013224}, which displays an unusual set of anomalies:  gate-tunable superconductivity with contrasting carrier‐density dependence of the SC transition temperature $T_c$ in weakly and strongly disordered regimes, and, most remarkably, a {\sl sudden} disappearance of SC  fluctuations below a critical doping. Such behaviors are in contrast to the mean-field BCS-theory  description, which, for ML systems, predicts superconductivity independent of the carrier density and of the disorder (Anderson theorem)~\cite{anderson1959theory,suhl1959impurity,skalski1964properties,andersen2020generalized}.

Physically, the reason why the ML SC systems are special is that they are fundamentally characterized by emerging SC-phase  fluctuations, in contrast to the bulk case where superconductivity is described solely by the SC gap within the mean-field theory. These fluctuations comprise the longitudinal Nambu–Goldstone (NG) phase fluctuations and the transverse Berezinskii–Kosterlitz–Thouless (BKT) fluctuations. 

The NG phase fluctuations arise as collective gapless long-wavelength modes dictated by the NG  theorem~\cite{nambu1960quasi,goldstone1961field,goldstone1962broken,nambu2009nobel}, following the spontaneous breaking of $U(1)$ symmetry in superconductors~\cite{ambegaokar1961electromagnetic,littlewood1981gauge}. Such smooth bosonic excitations~\cite{PhysRevB.62.6786,benfatto10,PhysRevB.64.140506,PhysRevB.69.184510,PhysRevB.70.214531,PhysRevB.97.054510,yang2021theory,benfatto2001phase} tend to restore the broken symmetry, thereby driving thermodynamic instabilities of the SC state.  Certain mechanisms can, in principle, circumvent this instability. In 3D superconductors, the NG mode is gapped by the Anderson–Higgs mechanism~\cite{anderson1963plasmons,ambegaokar1961electromagnetic,littlewood1981gauge,yang2019gauge,PhysRevB.97.054510} via long‐range Coulomb interactions, which shift the original gapless spectrum to high‐frequency plasma energy $\omega_p$, suppressing phase fluctuations entirely. The same mechanism renormalizing the NG‐mode dispersion to $\omega_p$ also operates in ML superconductors, as revealed by early~\cite{fisher1990presence,fisher1990quantum,mishonov1990plasmon,PhysRevB.69.184510} and recent~\cite{PhysRevB.97.054510,yang2021theory} studies on phase-mode spectrum. While the sublinear dispersion $\omega_p(q)\propto\sqrt{q}$ in ML is expected to avoid the infrared divergence of phase correlations at finite $T$~\cite{PhysRevB.97.054510}, providing a clear route to evade the Hohenberg–Mermin–Wagner–Coleman theorem~\cite{hohenberg1967existence,mermin1966absence,coleman1973there}, which forbids long-range SC order in ML systems, the gapless nature implies that the bosonic NG mode remains active and can play an essential role in renormalizing SC gap~\cite{PhysRevB.97.054510,yang2021theory,PhysRevB.70.214531,PhysRevB.102.060501}. 

The BKT fluctuations~\cite{doi:10.1142/9789814417648_0004,benfatto10,PhysRevB.110.144518} arise due to the discrete symmetry of phase transformation. These topological defects, which cannot be continuously deformed into the SC ground state, produce a universal discontinuous jump in the superfluid stiffness~\cite{PhysRevLett.131.186002,PhysRevB.100.064506},  driving the SC transition temperature $T_c$ below the gap‐closing temperature $T_{\mathrm{\rm os}}$ and creating a phase‐incoherent pairing regime above $T_c$~\cite{chand2012phase,mondal2011phase,PhysRevB.102.060501}. In this case, the SC-normal transition at $T_c$ is governed by the loss of long‐range phase coherence rather than by the closing of the pairing gap.  The standard BKT renormalization‐group (RG) approach requires the bare superfluid density as input, typically obtained from Ginzburg–Landau expression~\cite{Halperin1979,benfatto10} or the limiting form of BCS theory~\cite{PhysRevLett.131.186002,PhysRevB.102.060501}. This input may face challenges in disordered case~\cite{PhysRevB.107.224502,PhysRevB.99.104509,li2021superconductor,dubi2007nature}, since it does not account for the disorder effect owing to the disorder‐insensitivity predicted by the Anderson theorem~\cite{anderson1959theory,suhl1959impurity,skalski1964properties,andersen2020generalized}. In reality, ML materials, due to their reduced dimensionality, are intrinsically disordered, and experiments on disordered ultra-thin films~\cite{mondal2011phase,chockalingam2008superconducting,noat2013unconventional,chand2012phase,sacepe2008disorder,sacepe2020quantum,sacepe2011localization,sherman2015higgs,dubouchet2019collective} have reported that increasing the disorder suppresses $T_c$.

The ML SC system therefore simultaneously hosts fermionic quasiparticles, bosonic NG phase modes, BKT topological excitations, and disorder. Existing theoretical approaches usually address these ingredients separately in the limiting form, while current computational methods are optimized mainly for fermionic quasiparticles. Here, we combine these microscopic ingredients in a unified description, offering a self-consistent phase-transition theory for phase-fluctuating SC state. Specifically, starting from the purely microscopic model, we treat the SC gap by explicitly and self-consistently incorporating the bosonic excitation of the NG phase mode  with long‐range Coulomb interactions, and calculate the disorder‐modified superfluid density that enters the BKT RG equations. We show that when implemented with DFT-based parameters and excitonic competition, this theory quantitatively reproduces nearly all key experimental observations in gate-tuned superconductivity of  ML WTe$_2$ {in Refs.~\cite{fatemi2018electrically,sajadi2018gate,song2024unconventional,PhysRevResearch.7.013224}}, providing a consistent understanding of the reported anomalies.

{\sl SC model.---}{We begin with the standard microscopic Hamiltonian for $s$-wave superconductors 
\begin{eqnarray}
  &&H_0=\int\!\!{d{\bf x}}\Big[\sum_s\psi^{\dagger}_s({\bf x})\xi_{\bf {\hat p}}\psi_s({\bf x})-U\psi^{\dagger}_{\uparrow}({\bf x})\psi^{\dagger}_{\downarrow}({\bf x})\psi_{\downarrow}({\bf x})\psi_{\uparrow}({\bf x})\Big]\nonumber\\
  &&\mbox{}+\frac{1}{2}\sum_{ss'}\int{d{\bf x}d{\bf x}'}V({\bf x}-{\bf x}')\psi^{\dagger}_{s}({\bf x})\psi^{\dagger}_{s'}({\bf x}')\psi_{s'}({\bf x}')\psi_{s}({\bf x}).
\end{eqnarray}
Here, $\psi^{\dagger}_s({\bf x})$ and $\psi_s({\bf x})$ represent the creation and annihilation field operators of electron with spin $s=\uparrow,\downarrow$, respectively;  $U$ denotes the $s$-wave pairing potential; $\xi_{\bf {\hat p}}=\frac{{\bf {\hat p}^2}}{2m_e}-E_F$ with ${\hat {\bf p}}=-i\hbar{\bf \nabla}$ being the momentum operator and $E_F$ standing for the fermi energy; $V({\bf x-x'})$ represents the long-range Coulomb. Starting from this Hamiltonian, we formulate an effective action within the path-integral {(Sec.~SI~A)}, where the SC order parameter reads $\Delta=|\Delta|e^{i\delta\theta({\bf R})}$, 
with $|\Delta|$ and $\delta\theta({\bf R})$ being the gap and phase, respectively, and derive the theory {(Sec.~SI~B)} of SC gap and phase fluctuations using quantum-statistical approach. Here we only present a concise overview of the resulting framework, which can be directly implemented for practical calculations. {The full microscopic derivation is provided in Sec.~SI of the Supplementary Materials~\cite{supple} (including Refs.~\cite{yang2024diamagnetic,schrieffer1964theory,yang2023optical,abrikosov2012methods,PhysRevB.99.224511,PhysRevB.106.144509,eilenberger1968transformation,PhysRevLett.25.507,PhysRevB.76.094515,PhysRevB.72.064503,PhysRevB.10.1885,PhysRevB.6.2579,Landaubook,nam1967theory,mattis1958theory,verdi2023quantum,wu2022large,peskin2018introduction,muller1979srti,li2019terahertz,cheng2023terahertz,74d5-4hsw,PhysRevB.71.224502,Tang2023AmbipolarMoTe2,VASP1,VASP2,PBE,DFTD3,Xingen}). The detailed simulation treatments are addressed in Sec.~SIII.}  

 {The phase fluctuation is defined as ${\bf p}_s={\bf \nabla_R}\delta\theta({\bf R})/2$, and can be decomposed into two orthogonal components~\cite{benfatto10}:
 \begin{equation}
 {\bf p}_s={\bf p}_{s,\parallel}+{\bf p}_{s,\perp},~~\text{with}~~~{\bm \nabla}\times{\bf p}_{s,\parallel}=0~~~\text{and}~~~{\bm \nabla\cdot}{\bf p}_{s,\perp}=0.
\end{equation}
The longitudinal one ${\bf p}_{s,\parallel}$ (curl-free) is associated with NG long‐wavelength (smooth)  fluctuations, and the transverse one ${\bf p}_{s,\perp}$ (divergence-free) corresponds to the  BKT vortex fluctuations (topological defects) since only this part carries vorticity.}   The NG fluctuations, acting as a SC momentum~\cite{ambegaokar1961electromagnetic,nambu1960quasi,nambu2009nobel,littlewood1981gauge},  are expected to renormalize the SC gap~\cite{PhysRevB.97.054510,yang2021theory,PhysRevB.70.214531,PhysRevB.102.060501}, in a gauge manner analogous to the way a vector potential can influence the SC gap~\cite{ambegaokar1961electromagnetic,nambu1960quasi,nambu2009nobel}. The resulting gap equation {(Sec.~SI~B):}
\begin{equation}\label{GE}
 {\frac{1}{U}}=F({p}^2_{s,\parallel},|\Delta|,T)={\sum_{\bf k}}\frac{f(E_{\bf k}^+)-f(E_{\bf k}^-)}{2{E_{\bf k}}},
\end{equation}
where phase fluctuations enter through the Doppler shift ${\bf v_{\bf k}}\cdot{\bf p}_{s,\parallel}$~\cite{fulde1964superconductivity,larkin1965zh,yang2018fulde,yang2018gauge,yang2021theory} in the Bogoliubov quasiparticle energies {{$E_{\bf k}^{\pm}={\bf v_{\bf k}}\cdot{{\bf p_{s,\parallel}}}\pm{E_{\bf k}}$ with $E_{\bf k}=\sqrt{\xi_{\bf k}^2+|\Delta|^2}$}}. Here, $v_{\bf k}={\partial_{\bf k}}\xi_{\bf k}$ denotes the band velocity and $f(x)$ is the Fermi distribution function. The SC gap equation here is an even function of the phase fluctuations, and hence, influenced by the average {(Sec.~SI~B)}:
\begin{equation}\label{EPF}
{p}^2_{s,\parallel}=\langle{p_{s,{\rm NG}}^2}\rangle=\int\frac{d{\bf q}}{(2\pi)^2}\frac{q^2[2n_B(\omega_{\rm NG})+1]}{2D_{q}\omega_{\rm NG}(q)},  
\end{equation}
{which is a standard bosonic excitation $2n_B(\omega_{\rm NG})+1$ of the NG mode, and can naturally be classified into the thermal  excitations $2n_B(x)$ and zero-point oscillations~\cite{yang2021theory}}.  The energy spectrum of NG mode is derived as 
\begin{equation}
\omega_{\rm NG}(q)=\sqrt{n_sq^2/(2D_qm_e)},
\end{equation}
with {\small{$D_q=D/(1+2DV_q)$}}, as well established in the literature by various theoretical approaches~\cite{sun2020collective,yang2021theory,ambegaokar1961electromagnetic,yang2019gauge,PhysRevB.64.140506,PhysRevB.69.184510,PhysRevB.70.214531,PhysRevB.97.054510}. Here,  $D$ and {\small{$V_q=2{\pi}e^2/(q\epsilon_0)$}} denote the density of states of carriers and 2D Coulomb potential, respectively.

The superfluid density is obtained from the current–current correlation function, and written as {(Sec.~SI~B)}
\begin{equation}\label{SD}
\frac{n_s}{n}=\frac{1}{1+\xi/l}\int{d\xi_k}\int\frac{{d}\theta_{\bf k}}{2\pi}\frac{|\Delta|^2}{2E^3_{\bf k}}[f(E_{\bf k}^-)-f(E_{\bf k}^+)].  
\end{equation}
Here, a prefactor $(1+\xi/l)^{-1}$ is included to account for disorder effects, with $\xi=\hbar v_F/|\Delta|$ being the SC coherence length, $l=v_F\gamma^{-1}$ the mean free path, and $\gamma$ the effective  scattering rate, encompassing  
 the SC-phase-coherence dephasing time (rather than the momentum-relaxation time). This interpolation  was originally introduced by Tinkham in the context of the penetration depth of superconductors $\lambda$~\cite{tinkham2004introduction}, leading to $\lambda^2=\lambda_{\rm clean}^2(1+\xi/l)$ and hence $n_s\rightarrow n_s/(1+\xi/l)$. It captures the essential suppression of the superfluid density by nonmagnetic disorder, and shows that disorder affects the SC state only through the renormalization of $n_s$, without directly modifying the gap equation for $s$-wave pairing {(see Sec.~SI~C for detailed discussion)},  in accordance with Anderson's theorem~\cite{anderson1959theory,suhl1959impurity,skalski1964properties,andersen2020generalized}.

The bare superfluid density $n_s$ from Eq.~(\ref{SD}) incorporates the effects of fermionic quasiparticles, bosonic NG phase fluctuations, and disorder. Then, the standard BKT RG approach~\cite{benfatto10,PhysRevB.110.144518,PhysRevB.80.214506,PhysRevB.77.100506,PhysRevB.87.184505} is applied,
\begin{equation}
\frac{dK}{dl}=-K^2g^2~~~\text{and}~~~\frac{dg}{dl}=(2-K)g, \label{BKTEQ}
\end{equation}
with the initial condition $K(l=0)=\frac{\pi\hbar^2{n_s}}{4m_ek_BT}$ and $g(l=0)=2\pi{e}^{-cK(l=0)}$ where $c=2/\pi$ in the ML limit.  {Thus, the obtained bare $n_s$ naturally serves as the $l=0$ input to the BKT flow, 
and  with Eq.~(\ref{BKTEQ}),  integrating the BKT flow to $l\rightarrow\infty$ yields a renormalized superfluid density 
\begin{equation}
{\bar n}_s=\frac{4m_ek_BT}{\pi\hbar^2}K(l=\infty),
\end{equation}
which further accounts for the BKT fluctuations. The separatrix $2-\pi{K}=0$ gives the Nelson–Kosterlitz universal jump~\cite{benfatto10,PhysRevB.110.144518,PhysRevB.80.214506,PhysRevB.77.100506,PhysRevB.87.184505}, while flows with $K<2/\pi$ run to the disordered phase (unbound vortices, $g\rightarrow\infty$), and those with $K>2/\pi$ renormalize to a finite $K(l=\infty)$ with $g\rightarrow0$ (bound vortex–antivortex pairs)~\cite{benfatto10}.} A key quantity that represents the phase stiffness 
 is $n_s/m_e$, as established in numerous experiments~\cite{uemura1989universal,emery1995importance,uemura1991basic,yuli2008enhancement}, as its reduction lowers the excitation energy $\omega_{\rm NG}$ and coupling constant $K(l=0)$ and hence enhances both NG and BKT fluctuations.  Consequently, for each doping level (Fermi energy) and disorder strength, we self-consistently solve the gap equation, the NG phase fluctuations, and the disorder-modified bare superfluid density. Then, substituting the resulting bare finite-temperature superfluid density into the BKT renormalization-group equations allows us to determine the full set of SC  properties. The SC transition temperature $T_c$ is determined from the condition ${\bar n}_s(T_c)=0$ while the gap‐closing temperature $T_{\rm os}$ is obtained by $|\Delta(T=T_{\rm os})|=0$.

{\sl Excitonic instability.---}The ML WTe$_2$ typically crystallizes in  distorted 1$T'$-phase orthorhombic structure~\cite{doi:10.1126/science.1256815,Fei2017,Xu2018}. The band-structure calculations~\cite{doi:10.1126/science.1256815,Tang2017}  together with ARPES measurements~\cite{Tang2017} reveal that its Fermi surface consists of two electron pockets located near the $\pm{\bf Q}$ points along the $\Gamma$–$X$ direction and one hole pocket at the $\Gamma$ point, originating from band inversion between W $5d$ and Te $5p$ orbitals. The carrier density can be continuously tuned via electrostatic gating~\cite{doi:10.1126/science.aan6003,Jia2022,fatemi2018electrically,sajadi2018gate,song2024unconventional,PhysRevResearch.7.013224}.  However, due to the negative band gap $E_g<0$, hole states emerge once the Fermi level $E_F$ is lowered below $|E_g|$, and such holes can pair with electrons to form an excitonic insulator phase~\cite{Jia2022,PhysRev.158.462,PhysRevLett.19.439}, which competes with superconductivity. In ML WTe$_2$, the excitonic condensate is sufficiently strong to dominate over the electron SC channel~\cite{Jia2022,fatemi2018electrically,sajadi2018gate,song2024unconventional,PhysRevResearch.7.013224}.  {A fully microscopic treatment that simultaneously
describes both excitonic and SC orders beyond mean-field level requires a self-
consistent many-body calculation of intertwined phases, including fluctuation corrections in both sectors and a global minimization of the free energy to determine the energetically favored state. Such a comprehensive calculation of excitonic insulating order is not the focus of the present work for superconductivity, which is aimed at addressing how to go beyond mean-field theory from a microscopic perspective, as is fundamentally expected to be necessary in low-dimensional system.}

Therefore, to model this competition between two distinct many-body states, we assume that each hole binds with one electron to form an exciton (Sec. SIII), thereby depleting the electrons available for SC pairing as a consequence of pairing-electron-number conservation. The effective density of states entering the electron SC gap equation and the superfluid density is consequently reduced by a factor $(n_e-n_h)/n_e$. Here, $n_e=2n$ is the total conduction-band electron density and $n_h$ is the hole density,   both set by $E_F$ and $E_g$ in a parabolic-band 2DEG approximation. {This treatment of the excitonic sector should be understood as a controlled effective
approximation, included only as a competing ingredient to capture the experimentally observed suppression of superconductivity at low doping, rather than as a primary subject of superconductivity calculation, and is phenomenological in nature. However, it should also be emphasized that this instability only determines the critical doping for superconductivity in ML WTe$_2$. It is not a universal feature of 2D superconductors and does not alter the conclusions of this work concerning superconductivity}.

\begin{figure}[htb]
  {\includegraphics[width=8.7cm]{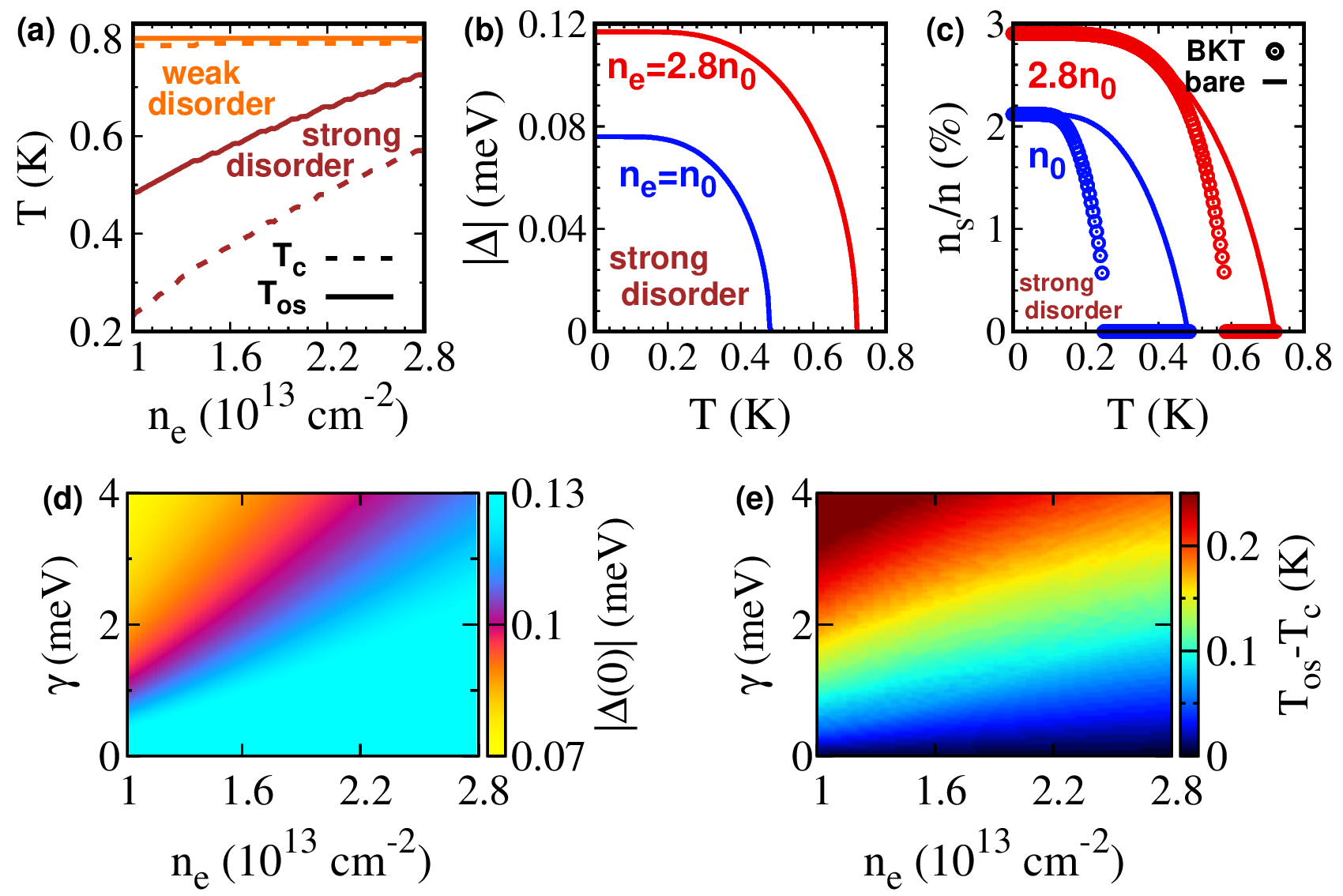}}
\caption{Simulation results of superconductivity in ML WTe$2$ at relatively high carrier density. ({\bf a}) $T_c$ (dashed) and $T_{\rm os}$ (solid) for weak disorder ($\gamma=0.12$ meV) and strong disorder ($\gamma=3.2$ meV). ({\bf b}) Temperature dependence of the SC gap under strong disorder ($\gamma=3.2$ meV) at two different densities. ({\bf c}) Bare (curve) and BKT-renormalized (circles) superfluid density corresponding to the results in panel ({\bf b}). ({\bf d}, {\bf e}) Zero-temperature gap $|\Delta(0)|$ and the temperature difference $T_{\rm os}-T_c$ versus density and disorder, respectively. The pairing potential is fixed and specified in the Supplemental Materials. }
\label{figyc1}
\end{figure}

{\sl Results.---}The effective masses of electrons and holes are obtained from our DFT calculations {(Sec.~SIV)} [Fig.~\ref{figyc2}(a)], but $E_g$ is sample dependent and sensitive to external conditions~\cite{PhysRevLett.125.046801}. The framework then involves three tunable parameters: {the band gap $E_g$ and the effective scattering rate $\gamma$ are fixed sample-dependent parameters, while the carrier density $n_e$ is tuned solely by adjusting the Fermi level $E_F$.} We first examine the density regime $n_e \in [1,2.8]\times10^{13}$ cm$^{-2}$, where holes are absent or negligible and superconductivity is therefore unaffected by the excitonic instability (set by $E_g$). 

In this density range and under weak disorder, $T_{\rm os}$ [orange solid curve in Fig.~\ref{figyc1}(a)] and the zero-temperature gap $|\Delta(0)|$ [small $\gamma$ in Fig.~\ref{figyc1}(d)] remain nearly density independent, so that the behavior essentially follows the mean-field description and NG fluctuations are negligible. This is consistent with the general conclusion that NG-fluctuation corrections to the gap are minimal in the clean limit~\cite{PhysRevB.70.214531}. The transition temperature $T_c$ [orange dashed curve in Fig.~\ref{figyc1}(a)] shows only a faint density dependence; consequently, the difference between $T_c$ and $T_{\rm os}$ is nonzero but small, as seen in the orange curves of Fig.~\ref{figyc1}(a) or in small-$\gamma$ case of Fig.~\ref{figyc1}(e), suggesting weak BKT fluctuations. The suppression of fluctuations originates from the small electron effective mass, $m_e = 0.33m_0$ (from our DFT calculations {(Sec.~SIV)}, consistent with the experimental estimate $m_e = 0.3m_0$~\cite{sajadi2018gate}), and the relatively high carrier density. Together, these yield a large phase stiffness $n_s/m_e$, which suppresses fluctuations.

In contrast, under strong disorder, where the superfluid density and thus the phase stiffness are strongly suppressed, fluctuations become significant. As shown in Fig.~\ref{figyc1}(c), BKT fluctuations at elevated temperatures cause the renormalized $\bar{n}_s$ to decrease more rapidly than the bare value, ultimately leading to a discontinuous drop of the superfluid density and driving the SC transition at $T_c$, with $T_c$ pushed significantly below the gap-closing temperature $T_{\rm os}$. This creates a pseudogap (phase-incoherent pairing) regime in the window $T_c < T < T_{\rm os}$, where pairing persists without the global phase coherence necessary to achieve superconductivity (zero-resistivity phenomenon). As a result, a large difference between $T_c$ and $T_{\rm os}$ emerges under strong disorder and becomes more pronounced as density (phase stiffness) decreases, as shown by brown curves in  Fig.~\ref{figyc1}(a), or in large-$\gamma$ case in Fig.~\ref{figyc1}(e). 

On the other hand, we find that the zero-temperature gap $|\Delta(0)|$ [Fig.~\ref{figyc1}(b),(d)], and consequently the gap-closing temperature $T_{\rm os}$ [brown solid curve in Fig.~\ref{figyc1}(a)] and the disorder-modified superfluid-density ratio $n_s(0)/n$ [Fig.~\ref{figyc1}(c)],  acquire a clear dependence on both density and disorder under strong disorder, decreasing as the density (phase stiffness) is reduced or disorder is enhanced. Such SC-gap behavior under strong disorder is fully consistent with the STM and transport measurements on ML NbSe$_2$~\cite{Zhao2019,sacepe2020quantum}. 
	 This emerging dependence arises from the  enhanced NG quantum fluctuations (bosonic zero-point oscillations) by disorder-suppressed phase stiffness, which couple back into the SC gap equation and renormalize $|\Delta(0)|$ in the strong-disorder case,  unlike the mean-field and conventional BKT descriptions in 2D case, where the SC gap remains insensitive to density variation and disorder. 

Nevertheless, we find that the NG thermal fluctuations remain minimal even under strong disorder, owing to the minimal thermal excitations of the NG mode $n_B(\omega_{\rm NG})$. This arises from the change of the  NG-mode energy spectrum from $\omega_{\rm NG}(q)\propto{q}$ to $\omega_{\rm NG}(q)\propto{\sqrt{q}}$ after accounting for the long-range Coulomb interactions, which confines low-energy excitations to a narrow momentum window with limited phase space. As a result, these modes are only weakly thermally populated, unlike linear-despersion modes, so their contributions to thermodynamics and critical dynamics become negligible. As proposed in Ref.~\cite{PhysRevB.97.054510,yang2021theory},  this behavior represents  an expected route to evade the infrared divergence of phase-fluctuation correlations at $T \neq 0$ imposed by the Mermin–Wagner  theorem~\cite{hohenberg1967existence,mermin1966absence,coleman1973there}. As a result, although the zero-temperature gap is renormalized by NG bosons in a density- and disorder-dependent manner [Fig.~\ref{figyc1}(d)], its temperature evolution nevertheless follows the BCS-theory description, as shown in Fig.~\ref{figyc1}(b).

  \begin{figure}[htb]
  {\includegraphics[width=8.6cm]{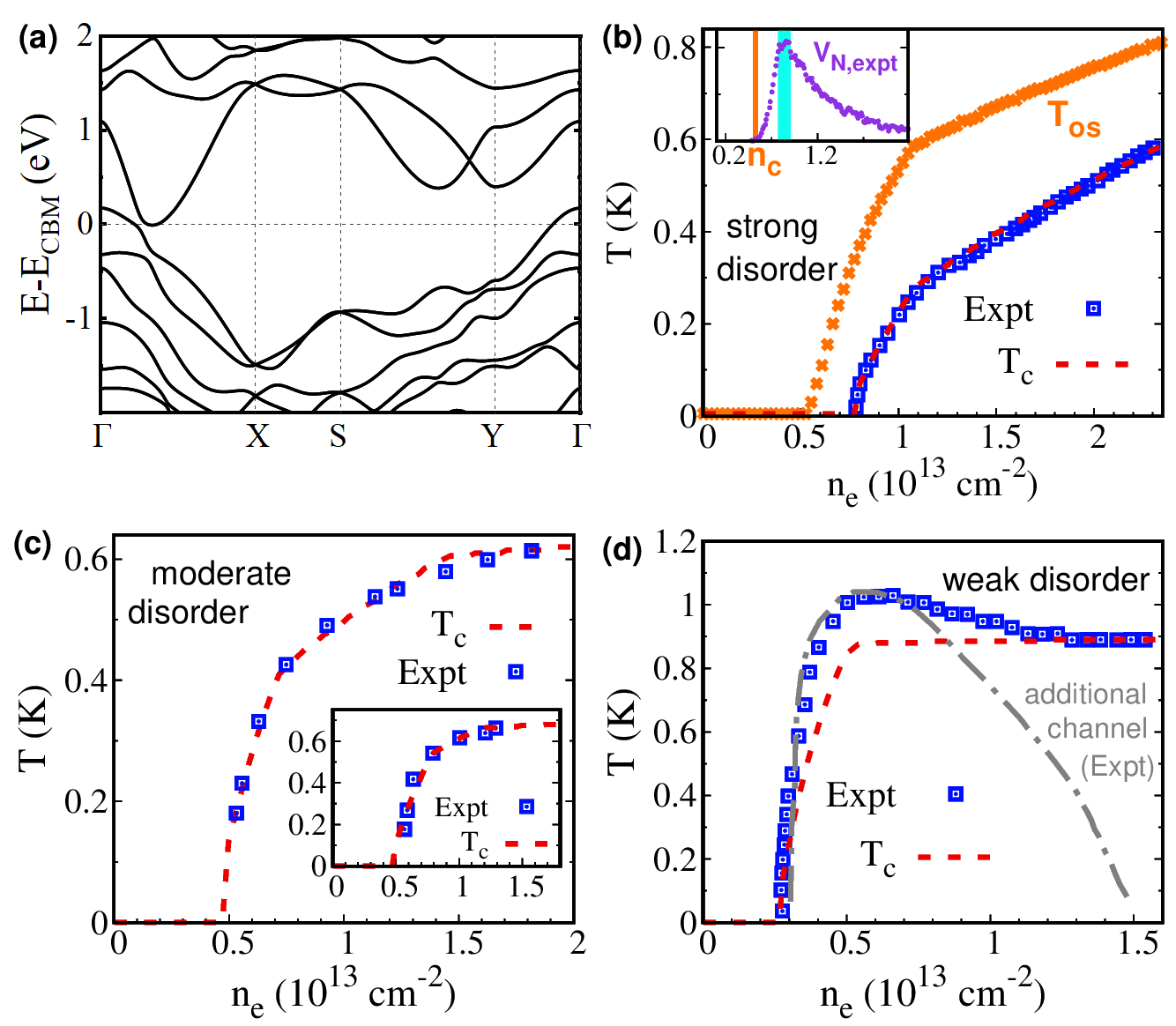}}
\caption{({\bf a}) DFT calculation of the band structure of ML WTe$_2$. ({\bf b}), ({\bf c}), inset of ({\bf c})~and~({\bf d}): experimentally measured $T_c$ from Refs.~\cite{song2024unconventional},~\cite{fatemi2018electrically},~\cite{sajadi2018gate}~and~\cite{PhysRevResearch.7.013224}, respectively, compared with our calculated $T_c$ (dashed curves) obtained by fitting $E_g$ and disorder strength. Crosses in ({\bf b}) denote our calculated $T_{\rm os}$. Inset of ({\bf b}): experimentally measured Nernst signal $V_n(T=45~\text{mK}, H=23~\text{mT})$ in Ref.~\cite{song2024unconventional}. The orange line marks our calculated $n_c$ where $T_{\rm os}$ vanishes, and the light blue region indicates the regime where our  $T_{\rm os}-T_c$ [Fig.~\ref{figyc3}({\bf b})] is maximized. The chain curve in ({\bf d}) is extracted from the experimentally observed superconductivity that survives at $H=50$ mT~\cite{PhysRevResearch.7.013224}. }

\label{figyc2}
  \end{figure}

After establishing the high-density behavior, we proceed to compare with experiments. The band gap $E_g$ is adjusted to reproduce the experimentally observed critical density (quantum critical point) for the onset of superconductivity, while the disorder strength $\gamma$ is determined from the high-density gap behavior discussed above. With these parameters fixed, as shown in Fig.~\ref{figyc2}, our framework quantitatively reproduces the experimentally observed density dependence of $T_c$ in ML WTe$_2$, and naturally accounts for the contrasting $T_c$–density trends observed across four different samples in the literature.

Specifically, as shown in {Fig.~\ref{figyc2}(b)}, the experimental data of Ref.~\cite{song2024unconventional} exhibit a pronounced density dependence of $T_c$ in the high-density regime, consistent with a strong-disorder scenario. Across the entire density range, our calculated $T_c(n)$ (dashed curve) nearly coincides with the experimental data (blue squares). Notably, the Nernst-signal measurements at a fixed magnetic field in Ref.~\cite{song2024unconventional} (purple dots in the inset of {Fig.~\ref{figyc2}(b)}) reveal a sudden disappearance of SC fluctuations below a critical density $n_c$. In our framework, this abrupt disappearance is explained by the vanishing of $T_{\rm os}$, signaling the termination of superconductivity by the excitonic instability, i.e., such disappearance of superconductivity at the critical doping $n_c$ is because the electronic states
participating in SC pairing are effectively exhausted, as the system crosses into an excitonic
insulating state that does not coexist with superconductivity. The density value $n_p$ at which the measured  Nernst signal peaks is associated with the density where $T_{\rm os}-T_c$ in our calculation is maximized [Fig.~\ref{figyc3}(b)], identifying the regime of strongest fluctuations. The measured  critical magnetic field $H_{c,n}$ [inset of Fig.~\ref{figyc3}(a)], at which the Nernst signal vanishes in the low-$T$ limit, corresponds to $H_{c2}(0)=\frac{4m^2_e|\Delta(0)|^2\Phi_0}{\pi^2\hbar^4n_s(0)}$ in our calculation {(Sec.~SII)}, above which Cooper pairs are fully suppressed. Remarkably, as shown in the inset of Fig.~\ref{figyc2}(b) and in Fig.~\ref{figyc3}(a), our predicted results, without adjustable parameters, accurately reproduce the experimental values of $n_c$, $n_p$, and, in particular, $H_{c,n}$, and its divergence-like dependence of as the density decreases~\cite{Discussion}, in a quantitative agreement.

  \begin{figure}[htb]
  {\includegraphics[width=8.6cm]{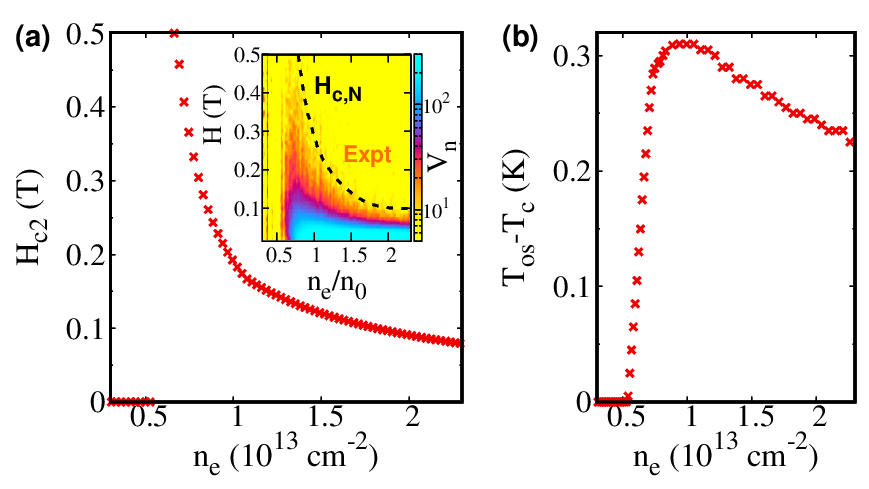}}
  \caption{Calculation of ({\bf a}) the critical field $H_{c2}(0)$ and ({\bf b}) the temperature dependence of $T_{\rm os}-T_c$ as a function of density, corresponding to simulation in Fig.~\ref{figyc2}({\bf b}). Inset of ({\bf a}): experimentally measured Nernst signal $V_n(T=45~\text{mK})$ (in unit of nV) from Ref.~\cite{song2024unconventional}; the dashed line indicates the field at which the signal vanishes in experiment.}    
\label{figyc3}
  \end{figure}

The experimental data of Ref.~\cite{fatemi2018electrically} [Fig.~\ref{figyc2}(c)] and Ref.~\cite{sajadi2018gate} [inset of Fig.~\ref{figyc2}(c)] exhibit a weak yet  visible density dependence of $T_c$ in the high-density range, consistent with a moderate-disorder scenario, and our calculated $T_c(n)$ closely follows the  experimental trend across entire density range.

The recently reported experimental data of Ref.~\cite{PhysRevResearch.7.013224} show only  minimal density variation of $T_c$ in the high-density range, consistent with a weak-disorder scenario. In this case, our results quantitatively capture the density independence of $T_c$ at high densities ($n_e>1\times10^{13}/\text{cm}^2$) and qualitatively describes its rapid suppression as the excitonic instability is approached ($n_e<0.5\times10^{13}/~\text{cm}^2$), as shown in Fig.~\ref{figyc2}(d). However, in this sample a small anomalous upturn of $T_c$ appears within the density range $n_e\in(0.5,1)\times10^{13}~\text{cm}^{-2}$, where $T_c$ increases upon reducing $n_e$, deviating from our theoretical curve.

Experimental observations of this anomalous upturn remain scarce, so far limited to the study in Ref.~\cite{PhysRevResearch.7.013224}. While it could be a measurement artifact (e.g., electrostatic screening or device geometry), it may also suggest the presence of an additional SC  pairing channel active in this density window~\cite{
doi:10.1073/pnas.2117735119}.  
 {In the experiment of Ref.~\cite{PhysRevResearch.7.013224}, upon gradually applying an external magnetic field that selectively suppresses the background SC component, with the suppression setting in first on the high-density side, the visibility of  the low-density SC behavior is enhanced. As a result, the residual $T_c$ exhibits a  distinct $T_c$–$n_e$ dome behavior, as indicated by the gray chain curve in Fig.~\ref{figyc2}(d) (extracted from the 50~mT  data), and this dome sharpens at larger fields and persists up to $\sim$250~mT. In this case, as seen in Fig.~\ref{figyc2}(d),  combining this field-robust additional-channel $T_c$ with our calculated background $T_c$ (red dashed curve),  i.e., taking the larger of the two (short-circuit effect), reproduces the experimentally measured  zero-field $T_c(n)$ (blue squares).}

{This channel may stem from a hole-sector SC contribution on the electron-doped side, yielding a multiband superconductivity akin to MgB$_2$. Owing to the band inversion in ML WTe$_2$, holes and electrons share similar orbital character on the electron-doped side, implying comparable pairing interactions and thus the potential for hole-SC state.  The larger DOS of holes ($m_h=0.53m_0>m_e$) suggests that hole-SC pairing gap can be strong and may transiently overcome the excitonic competition in this density window. On the electron-doping side, the hole concentration remains low. This combination (low $n_h$, large $m_h$, and a sizable SC gap) implies a shorter SC coherence length, making hole-SC state more robust against magnetic-field depairing, with $H_{c2}^h$ expected to exceed  $H_{c2}^e$.  But this additional SC  channel inferred on the electron-doped side is unlikely on the hole-doped side, where the orbital character differs and the absence of band inversion likely precludes the same pairing mechanism from forming~\cite{PhysRevLett.125.237006}.}

{A complete understanding requires additional evidences to determine whether other candidate SC channel  exists. Accordingly, we confine analysis to the background component and related anomalies. Within this scope, our framework, which explicitly incorporates the relevant microscopic ingredients (fermionic quasiparticles, bosonic phase dynamics, topological defects, and disorder) beyond mean-field theory, quantitatively reproduces nearly all key experimental observations of gate-tunable superconductivity in ML WTe$_2$. This theoretical framework and the resulting findings should be general, as they can be broadly applicable not only to WTe$_2$ but also to other ML SC materials discovered over the past few decades as well as the rapidly growing family of 2D SC systems.}

{Finally, while the present work includes the orbital effects of a magnetic field and focuses on the zero-Zeeman-field regime, we note that further incorporating Zeeman-dependent modifications of the quasiparticle spectrum and phase fluctuations, such as those arising from spin–orbit–parity coupling~\cite{PhysRevLett.125.107001}, provides a natural route to extend the present approach to superconductivity at in-plane magnetic fields  (where orbital effects are negligible and Zeeman physics dominates) above the Pauli limit in gate-induced ML WTe$_2$.}

{\it Acknowledgments.---}
  This work is supported by the US Department of Energy, Office of Science, Basic Energy Sciences, under Award Number DE-SC0020145 as part of Computational Materials Sciences Program.  F.Y. and L.Q.C. also appreciate the generous support from the Donald W. Hamer Foundation through a Hamer Professorship at Penn State.


\begin{thebibliography}{127}
\bibitem [{Dis()}]{Discussion}%
  \BibitemOpen
  \href@noop {} {}\bibinfo {note} {This divergence-like density dependence of
  the critical field $H_{c,n}(0)$ reported in
  Ref.~\cite{PhysRevResearch.7.013224} was attributed to an unconventional
  pairing channel, where the resulting SC gap $|\Delta(0)|$ is assumed to
  increase at lower densities while $T_c$ is suppressed, thereby implying a
  decoupling between the $T_c$ and gap behaviors. This interpretation, however,
  relies on the assumption that superconductivity is destroyed solely by the
  Zeeman depairing effect of the magnetic field, i.e., $H_{c,n}(0)\propto
  |\Delta(0)|$. Alternatively, our analysis suggests that in Nernst
  measurements the relevant threshold is the orbital depairing of magnetic
  field, yielding the scaling relation $H_{c,n}(0)\propto
  |\Delta(0)|/n_s(0)\propto|\Delta(0)|(1+\xi/l)/(2n_e-n_h)$. This offers a
  natural explanation of the apparent divergence as approaching the critical
  density without requiring an unconventional pairing mechanism.}\BibitemShut
  {Stop}%
\end{thebibliography}
%

\end{document}


\preprint{APS/123-QED}

\title{Microscopic Phase-Transition Framework for Gate-Tunable Superconductivity in Monolayer WTe$_2$ (Supplementary Materials)}

\author{F. Yang}
\email{fzy5099@psu.edu}

\affiliation{Department of Materials Science and Engineering and Materials Research Institute, The Pennsylvania State University, University Park, PA 16802, USA}

\author{G. D. Zhao}
\affiliation{Department of Materials Science and Engineering and Materials Research Institute, The Pennsylvania State University, University Park, PA 16802, USA}

\author{Y. Shi}
\affiliation{Department of Materials Science and Engineering and Materials Research Institute, The Pennsylvania State University, University Park, PA 16802, USA}

\author{L. Q. Chen}
\email{lqc3@psu.edu}

\affiliation{Department of Materials Science and Engineering and Materials Research Institute, The Pennsylvania State University, University Park, PA 16802, USA}

\date{\today}

\maketitle

\section{Derivation of theoretical framework}

The theoretical framework of the present work has been summarized in the main text. {Detailed simulation treatments of this framework are addressed in Sec.~\ref{secst}.} In this section, we introduce its fundamental derivation from the microscopic superconducting (SC) model within the path-integral
and quantum-statistical approach.\\

{\bf  A general description of the steps}: We start with the standard microscopic Hamiltonian of the $s$-wave superconductors~\cite{li2021superconductor,dubi2007nature,yang2019gauge,yang2021theory}:
\begin{equation}\label{Ham}
  H_0=\int\!\!{d{\bf x}}\Big[\sum_s\psi^{\dagger}_s({\bf x})\xi_{\bf {\hat p}}\psi_s({\bf x})-U\psi^{\dagger}_{\uparrow}({\bf x})\psi^{\dagger}_{\downarrow}({\bf x})\psi_{\downarrow}({\bf x})\psi_{\uparrow}({\bf x})\Big]+\frac{1}{2}\sum_{ss'}\int{d{\bf x}d{\bf x}'}V({\bf x}-{\bf x}')\psi^{\dagger}_{s}({\bf x})\psi^{\dagger}_{s'}({\bf x}')\psi_{s'}({\bf x}')\psi_{s}({\bf x}).
\end{equation}
Here, $\psi^{\dagger}_s({\bf x})$ and $\psi_s({\bf x})$ represent the creation and annihilation field operators of electron with spin $s=\uparrow,\downarrow$, respectively;  $U$ denotes the $s$-wave pairing potential; $\xi_{\bf {\hat p}}=\frac{{\bf {\hat p}^2}}{2m}-E_F$ with ${\hat {\bf p}}=-i\hbar{\bf \nabla}$ being the momentum operator and $E_F$ standing for the fermi energy; $V({\bf x-x'})$ represents the long-range Coulomb.  From the Hamiltonian, one can write down the action of this model and apply the Hubbard-Stratonovich transformation to obtain the action of superconductors. Then, by separating the center-of-mass $[R=\frac{x_1+x_2}{2}=(t,{\bf R})]$ and relative $[r=x_1-x_2=(\tau,{\bf r})]$ coordinates of the pairing electrons and following the standard treatment within the path-integral approach~\cite{schrieffer1964theory,yang2021theory,yang2024diamagnetic,yang2023optical,sun2020collective}, the effective action of the gap and phase fluctuation is given by (the detailed  derivation is addressed in Sec.~\ref{EA})
\begin{eqnarray}\label{action}
{S}_{\rm eff}=\int{d{R}}\Big\{{\sum_{p_n,{\bf k}}}\ln[(ip_n-E_{\bf k}^+)(ip_n-E_{\bf k}^-)]\!-\!\frac{|\Delta|^2}{U}-\frac{np_s^2}{2m}\Big\}+\!\!\int{dt}\sum_{\bf q}\frac{D}{1\!+\!2DV_q}\Big(\frac{\partial_{t}\delta\theta_{\bf q}}{2}\!\Big)^2.
\end{eqnarray}
Here, the Doppler-shifted quasiparticle energy spectra {{$E_{\bf k}^{\pm}={\bf v_{\bf k}}\cdot{{\bf p_s}}\pm{E_{\bf k}}$}} with $E_{\bf k}=\sqrt{\xi_{\bf k}^2+|\Delta|^2}$; $p_n=(2n+1)\pi{T}$ denotes the Fermion Matsubara frequency; ${\bf k}$ and ${\bf q}$ represent the momenta of Fermion and Boson that are transferred from the relative and center-of-mass spatial coordinates, respectively. 

Through the variation $\delta{S_{\rm eff}}$ with respect to the gap, the gap equation is obtained (Sec.~\ref{EB}):
\begin{equation}\label{GE}
 {\frac{1}{U}}=F({p}^2_{s},|\Delta|,T)={\sum_{\bf k}}\frac{f(E_{\bf k}^+)-f(E_{\bf k}^-)}{2{E_{\bf k}}}.
\end{equation}
Equation~\eqref{GE} has the same structure as in Fulde–Ferrell–Larkin–Ovchinnikov–type treatments with finite center-of-mass  momentum~\cite{fulde1964superconductivity,larkin1965zh,yang2018fulde,yang2018gauge,yang2021theory}, and by particle-hole and angular averaging symmetries, the dependence enters as $F({p}^2_{s},|\Delta|,T)$.

Accordingly, the phase-fluctuation field ${\bf p}_s={\bf \nabla_R}\delta\theta({\bf R})/2$ emerges in the SC action and gap equation here. Following discussion in Ref.~\cite{benfatto10}, the phase dynamics ${\bf p}_{s}$ can be decomposed into two orthogonal components (Helmholtz decomposition)~\cite{benfatto10}: 
 \begin{equation}
 {\bf p}_s={\bf p}_{s,\parallel}+{\bf p}_{s,\perp},~~\text{with}~~~{\bm \nabla}\times{\bf p}_{s,\parallel}=0~~~\text{and}~~~{\bm \nabla\cdot}{\bf p}_{s,\perp}=0.
\end{equation} 

The longitudinal one ${\bf p}_{s,\parallel}$ 
is associated with the gapless Nambu–Goldstone (NG) smooth, long-wavelength phase fluctuations and acts as the SC momentum~\cite{ambegaokar1961electromagnetic,nambu1960quasi,nambu2009nobel,littlewood1981gauge}. Specifically, through the variation $\delta{S_{\rm eff}}$ with respect to the phase fluctuation and considering the long-wavelength approximation, one can find the equation of motion of ${\bf p}_{s,\parallel}$:
\begin{equation}\label{EOMPS}
[\partial_t^2+\omega^2_{\rm NG}(q)]{\bf p}_{s,\parallel}({\bf q})=0,  
\end{equation}
where the energy spectrum of NG mode is given by $\omega^2_{\rm NG}(q)=n_sq^2/(2D_qm)$ and the superfluid density is written as (Sec.~\ref{EB})
\begin{equation}\label{SD}
n_s=2E_F|\Delta|^2\sum_{\bf k}\frac{\partial_{E_{\bf k}}}{E_{\bf k}}\Big[\frac{f(E_{\bf k}^+)-f(E_{\bf k}^-)}{2E_{\bf k}}\Big]\approx2E_F|\Delta|^2\sum_{\bf k}\frac{f(E_{\bf k}^-)-f(E_{\bf k}^+)}{2E^3_{\bf k}}.  
\end{equation}
It is noted that Eq.~(\ref{EOMPS}) is exactly same as the equation of motion of the NG mode in the literature~\cite{ambegaokar1961electromagnetic,nambu1960quasi,nambu2009nobel,littlewood1981gauge,yang2021theory,sun2020collective} and the expression of the superfluid density is also same as the one obtained in the literature by various approaches~\cite{yang2021theory,sun2020collective,yang2018gauge,yang2019gauge,yang2024diamagnetic,abrikosov2012methods,ambegaokar1961electromagnetic}. The statistical average of these fluctuations must be treated within quantum-statistical framework, as a consequence of the NG bosons dictated by fundamental NG  theorem~\cite{nambu1960quasi,goldstone1961field,goldstone1962broken,nambu2009nobel}, following the spontaneous breaking of $U(1)$ symmetry in superconductors.  From the equation of motion of NG mode, one can introduce a thermal field and then calculate the thermal phase fluctuation via the fluctuation dissipation theorem, or directly calculate the thermal phase fluctuation through the Bosonic Green function within Matsubara representation. Both approaches lead to the same results: $\langle{\bf p}_{s,\parallel}\rangle=0$ and
\begin{equation}\label{BET}
\langle{p_{s,\parallel}^2}\rangle=S_{\rm th}(T)+S_{\rm zo},
\end{equation}
with the thermal-excitation part
\begin{equation}\label{th}
S_{\rm th}(T)=\int\frac{d{\bf q}}{(2\pi)^2}\frac{q^2n_B(\omega_{\rm NG})}{D_{q}\omega_{\rm NG}(q)}, 
\end{equation}
and the zero-point oscillation part 
\begin{equation}\label{zo}
S_{\rm zo}=\int\frac{d{\bf q}}{(2\pi)^2}\frac{q^2}{2D_{q}\omega_{\rm NG}(q)}.  
\end{equation}
Here, $n_B(x)$ denotes the Bose-Einstein distribution.
 This part enters the SC gap equation and is expected to renormalize the SC gap~\cite{PhysRevB.97.054510,yang2021theory,PhysRevB.70.214531,PhysRevB.102.060501}, in a gauge manner analogous to the way a vector potential can influence the SC gap~\cite{ambegaokar1961electromagnetic,nambu1960quasi,nambu2009nobel}. 

The transverse one ${\bf p}_{s,\perp}$ encodes Berezinskii–Kosterlitz–Thouless (BKT) fluctuations, since only this part carries vorticity: 
\begin{equation}
\pi\sum_jq_j=\oint{\bf p}_s\cdot{d{\bf l}}=\int_S({\bm \nabla}\times{\bf p_s})\cdot{d{\bf s}}=\int_S({\bm \nabla}\times{\bf p_{s,\perp}})\cdot{d{\bf s}},
\end{equation}
where $j$ indexes individual vortices or antivortices and $q_j\in\mathbb{Z}$ is their integer vorticity (topological charge). Equivalently, ${\bm \nabla}\times{\bf p_{s,\perp}}=\pi\sum_jq_j\delta({\bf r-r_j})$, a sum of point-like topological defects located at position ${\bf r_j}$. These topological defects (vortices) primarily disorder the SC phase and renormalize the superfluid stiffness, while the pairing amplitude remains essentially unchanged except within the vortex-core singularities.  The BKT renormalization‐group (RG) description  is well established in the literature, and takes the bare superfluid density $n_s$ as input to the standard RG equations~\cite{benfatto10,PhysRevB.110.144518,PhysRevB.80.214506,PhysRevB.77.100506,PhysRevB.87.184505}:
\begin{equation}\label{BKT1}
\frac{dK}{dl}=-K^2g^2~~~\text{and}~~~\frac{dg}{dl}=(2-K)g,
\end{equation}
in which the initial conditions of the dimensionless stiffness $K(l)$ and the vortex fugacity $g(l)$ are given by 
\begin{equation}
K(l=0)=\frac{\pi}{k_BT}\frac{\hbar^2{n_s}}{4m}~~~\text{and}~~~g(l=0)=2\pi{e}^{-\mu_v(T)/(k_BT)}
\end{equation}
Here, $\mu_v$ is the vortex–core energy, and  following Ref.~\cite{benfatto10} for 3D superconductors, in the monolayer limit, the vortex–core energy may be estimated as the condensation energy lost inside a core of radius $L_v$: 
\begin{equation}\label{BKT2}
\mu_v=\pi{L^2_v}E_{c}\frac{n_s}{n}=\pi\frac{\hbar^2v_F^2}{\pi^2|\Delta|^2}\frac{D|\Delta|^2}{2}\frac{n_s}{n}=\frac{2}{\pi}\frac{\hbar^2n_s}{4m}=\frac{2}{\pi^2}\frac{\pi\hbar^2n_s}{4m}, 
\end{equation}
with $L_v=\hbar{v_F}/(\pi|\Delta|)$ and $E_{c}={D|\Delta|^2}/{2}$ being the condensation-energy density. Thus, with Eqs.~(\ref{BKT1})-(\ref{BKT2}),  using the bare superfluid density, integrating the BKT flow to $l\rightarrow\infty$ yields a renormalized value ${\bar n}_s=\frac{4m_ek_BT}{\pi\hbar^2}K(l=\infty)$, which accounts for the vortex-antivortex fluctuations. The separatrix $2-\pi{K}=0$ gives the Nelson–Kosterlitz universal jump~\cite{benfatto10,PhysRevB.110.144518,PhysRevB.80.214506,PhysRevB.77.100506,PhysRevB.87.184505}, while flows with $K<2/\pi$ run to the incoherent phase (unbound vortices, $g\rightarrow\infty$) and those with $K>2/\pi$ renormalize to a finite $K(l=\infty)$ with $g\rightarrow0$ (bound vortex–antivortex pairs)~\cite{benfatto10}.  

Consequently, consistent with the longitudinal and transverse decomposition, only the longitudinal (curl-free) component, NG phase fluctuations, enters the SC gap equation and drives uniform pair breaking; the transverse (divergence-free) component contributes via BKT vortex physics to the renormalization of the superfluid stiffness but does not directly modify the homogeneous amplitude equation, except for local suppression inside vortex cores. 

{\sl Disorder Modification.---}It should be emphasized that the above derivation holds in the clean limit. In practice, disorder is unavoidable in monolayer systems and affects the amplitude (gap) and phase (stiffness) sectors differently. For isotropic
 $s$-wave pairing, the SC gap equation is not renormalized by nonmagnetic impurities (Anderson theorem~\cite{anderson1959theory,suhl1959impurity,skalski1964properties,andersen2020generalized}, Sec.~\ref{ED}), whereas the superfluid density, being a current–current correlation~\cite{PhysRevB.99.224511,PhysRevB.106.144509}, is reduced by the elastic scattering, as realized by various theoretical approaches using Gorkov theory~\cite{abrikosov2012methods}, Eilenberger transport~\cite{eilenberger1968transformation,PhysRevLett.25.507,PhysRevB.99.224511}, gauge-invariant kinetic equations~\cite{yang2018gauge}, and diagrammatic formulations incorporating {\sl vertex} corrections to the current–current correlation~\cite{PhysRevB.99.224511,PhysRevB.106.144509}. While illuminating, such microscopic treatments are cumbersome for practical modeling to compare with experiments. Here we directly adopt Tinkham's interpolation in the context of $s$-wave superconductors~\cite{tinkham2004introduction}, 
 which approximate the penetration depth $\lambda$ as 
 $\lambda^2=\lambda_{\rm clean}^2(1+\xi/l)$ and hence $n_s\rightarrow n_{s,{\rm clean}}/(1+\xi/l)$. Thus, the superfluid-density expression [Eq.~(\ref{SD})] in our framework becomes 
 \begin{equation}\label{SD2}
\frac{n_s}{n}=\frac{1}{1+\xi/l}\int{d\xi_k}\int\frac{{d}\theta_{\bf k}}{2\pi}\frac{|\Delta|^2}{2E^3_{\bf k}}[f(E_{\bf k}^-)-f(E_{\bf k}^+)],  
\end{equation}
 where $\xi=\hbar v_F/|\Delta|$ is the SC coherence length and $l=v_F\gamma^{-1}$ is the mean free path, with $\gamma$ representing the effective  scattering rate, encompassing  
 the SC-phase-coherence dephasing time and implicitly including the localization effects~\cite{li2021superconductor,dubi2007nature,sacepe2008disorder,sacepe2020quantum,sacepe2011localization,sherman2015higgs,dubouchet2019collective} in the disordered regime. This prescription has been shown to successfully account for early experimental measurements of the penetration depth $\lambda$~\cite{PhysRevB.76.094515,PhysRevB.72.064503,PhysRevB.10.1885,PhysRevB.6.2579}.\\

In the following, we present the detailed length derivations.

\subsection{Derivation of effective action}
\label{EA}

From the Hamiltonian (\ref{Ham}), the action of this model is given by~\cite{schrieffer1964theory,yang2021theory,yang2024diamagnetic,yang2023optical,sun2020collective} 
\begin{equation}
  S[\psi,\psi^{\dagger}]=\int{dx}\Big\{\sum_{s}\psi^{\dagger}_s(x)(i\partial_{x_0}-\xi_{\hat {\bf p}})\psi_s(x)+U\psi^{\dagger}_{\uparrow}(x)\psi^{\dagger}_{\downarrow}(x)\psi_{\downarrow}(x)\psi_{\uparrow}(x)\Big\}\!-\!\frac{1}{2}\sum_{ss'}\int{dxdx'}V(x-x')\psi_s^{\dagger}(x)\psi_{s'}^{\dagger}(x')\psi_{s'}(x')\psi_s(x),
\end{equation}
where the four-vector ${x}=(x_0,{\bf x})$.  Applying the Hubbard-Stratonovich transformation~\cite{schrieffer1964theory,yang2021theory,yang2024diamagnetic,yang2023optical,sun2020collective}, one has 
\begin{eqnarray}
S[\psi,\psi^{\dagger}]=\sum_{s}\int\!\!\!{dx}\psi^{\dagger}_s(x)[i\partial_{x_0}\!-\!\xi_{\hat {\bf p}}\!-\!\mu_H(x)]\psi_s(x)\!-\!\!\int\!\!\!{dx}\Big[\psi^{\dagger}(x){\hat \Delta}(x)\psi(x)\!+\!\frac{|\Delta(x)|^2}{U}\Big]\!+\!\frac{1}{2}\int\!\!\!{dx_0}d{\bf q}\frac{|\mu_H(x_0,{\bf q})|^2}{V_{\bf q}}.
\end{eqnarray}
Here, $\psi(x)=[\psi_{\uparrow}(x),\psi^{\dagger}_{\downarrow}(x)]^T$ represents the field operator in Nambu space; ${\hat \Delta}(x)=\Delta(x)\tau_++\Delta^*(x)\tau_-$ with $\tau_i$ being the Pauli matrices in Nambu space~\cite{abrikosov2012methods}; the  SC order parameter $\Delta(x)=|\Delta(x)|e^{i\delta\theta(x)}$ with $|\Delta(x)|$ and $\delta\theta(x)$ being the SC gap and phase, respectively;  $\mu_H(x_0,{\bf q})$ stands for the Hartree field.  

Using the unitary transformation $\psi(x){\rightarrow}e^{i\tau_3\delta\theta(x)/2}\psi(x)$ to effectively remove the SC  phase from the order parameter, the above action becomes
\begin{equation}
S=\!\!\int\!\!\!{dx}\Big\{\sum_{s}\psi^{\dagger}_s(x)\Big[i\partial_{x_0}-\frac{\partial_{x_0}\delta\theta(x)}{2}-\xi_{{\hat {\bf p}}+{\bf p}_s}-\mu_H(x)\Big]\psi_s(x)-\psi^{\dagger}(x)|{\Delta}(x)|\tau_1\psi(x)-\frac{|\Delta(x)|^2}{U}\Big\}+\int\!\!\!{dx_0}d{\bf q}\frac{|\mu_H(x_0,{\bf q})|^2}{2V_{\bf q}},
\end{equation}
where the phase fluctuations ${\bf p}_s={\bf \nabla}\delta\theta/2$. 

By separating center-of-mass $[R=\frac{x_1+x_2}{2}=(t,{\bf R})]$ and relative $[r=x_1-x_2=(\tau,{\bf r})]$ coordinates of the pairing electrons~\cite{schrieffer1964theory,yang2021theory,yang2024diamagnetic,yang2023optical,sun2020collective,yang2019gauge,abrikosov2012methods}, within the assumption of a macroscopically homogeneous gap and $\xi_{{\hat {\bf p}}+{\bf p}_s}=\xi_{\bf {\hat k}}+{\bf p}_s\cdot{\bf v_{\hat k}}+{\bf p}_s^2/(2m)$, one obtains
\begin{eqnarray}
  S[\psi,\psi^{\dagger}]&=&\int{dx}\psi^{\dagger}(x)\Big\{(i\partial_{\tau}\!-\!{\bf p}_s\cdot{\bf v_{\bf {\hat k}}}\!-\!\xi_{\hat {\bf k}}\tau_3\!-\!|\Delta|\tau_1)\!+\!\Big[\frac{p_s^2}{2m}\!+\!\mu_H({R})\!+\!\frac{\partial_t\delta\theta(R)}{2}\Big]\tau_3\Big\}\psi(x)\nonumber\\
  &&\mbox{}-\!\int{dR}\Big(\frac{p_s^2}{2m}\!+\!\frac{|\Delta|^2}{U}\!+\!\mu_H(R)\!+\!\frac{\partial_t\delta\theta(R)}{2}\Big)\!+\!\frac{1}{2}\int{dt}d{\bf q}\frac{|\mu_H({\bf q},x_0)|^2}{V_q}\nonumber\\
  &=&\int{dx}\psi^{\dagger}(x)\Big\{(i\partial_{\tau}-{\bf p}_s\cdot{\bf v_{\bf {\hat k}}}-\xi_{\hat {\bf k}}\tau_3-|\Delta|\tau_1)+\Big[\frac{p_s^2}{2m}+\mu_H({R})+\frac{\partial_t\delta\theta(R)}{2}\Big]\tau_3\Big\}\psi(x)\nonumber\\
  &&\mbox{}-\!\int{dR}\Big(\frac{p_s^2}{2m}+\frac{|\Delta|^2}{U}\Big)+\!\frac{1}{2}\int{dt}d{\bf q}\frac{|\mu_H({\bf q},x_0)|^2}{V_q}.
\end{eqnarray}
Then, through the standard integration over the Fermi field~\cite{schrieffer1964theory,yang2021theory,yang2024diamagnetic,yang2023optical,sun2020collective}, one has
\begin{eqnarray}
S=\int{dR}\Big({\rm {\bar T}r}\ln{G_0^{-1}}-\sum_n^{\infty}\frac{1}{n}{\rm {\bar T}r}\{[\Sigma(R)G_0]^n\}\Big)-\int{dR}\Big(\frac{p_s^2}{2m}+\frac{|\Delta|^2}{U}\Big)+\frac{1}{2}\int{dt}d{\bf q}\frac{|\mu_H({\bf q},x_0)|^2}{V_q},
\end{eqnarray}  
where the Green function in the Matsubara representation and momentum space: 
\begin{equation}\label{GreenfunctionMr}
G_0(p)=\frac{ip_n\tau_0-{\bf p}_s\cdot{\bf v_k}\tau_0+\xi_{\bf k}\tau_3+|\Delta|\tau_1}{(ip_n-E_{\bf k}^+)(ip_n-E_{\bf k}^-)},  
\end{equation}  
and self-energy: 
\begin{equation}
\Sigma=\Big[\frac{p_s^2}{2m}+\mu_H(R)+\frac{\partial_t\delta\theta(R)}{2}\Big]\tau_3.  
\end{equation}
Here, the four-vector $p=(ip_n, {\bf k})$.  Keeping the lowest two orders (i.e., $n=1$ and $n=2$) leads to the result:
\begin{equation}
  S=\int{dR}\Big\{\sum_{p_n,{\bf k}}\ln[(ip_n-E_{\bf k}^+)(ip_n-E_{\bf k}^-)]-{\tilde \chi}_3\frac{p_s^2}{2m}+\chi_{33}\Big[\frac{p_s^2}{2m}+\mu_H({R})+\frac{\partial_t\delta\theta(R)}{2}\Big]^2\!-\!\frac{|\Delta|^2}{U}\Big\}+\frac{1}{2}\int{dt}d{\bf q}\frac{|\mu_H({\bf q},x_0)|^2}{V_q},
\end{equation}  
where the correlation coefficients are given by
\begin{eqnarray}
{\tilde \chi}_3&=&\sum_{\bf k}\Big[1+\sum_{p_n}{\rm Tr}[G_0(p)\tau_3]\Big]=\sum_{\bf k}\Big[1+\frac{\xi_{\bf k}}{E_{\bf k}}\Big(f(E_{\bf k}^+)-f(E_{\bf k}^-)\Big)\Big]=-\frac{k_F^2}{2m}\sum_{\bf k}\partial_{\xi_{\bf k}}\Big[\frac{\xi_{\bf k}}{E_{\bf k}}\Big(f(E_{\bf k}^+)-f(E_{\bf k}^-)\Big)\Big]=n, \\
\chi_{33}&=&-\frac{1}{2}\sum_p{\rm Tr}[G_0(p)\tau_3G_0(p)\tau_3]=-\sum_{p_n,{\bf k}}\frac{(ip_n-{\bf p}_s\cdot{\bf v_k})^2+\xi_{\bf k}^2-|\Delta|^2}{(ip_n-E_{\bf k}^+)^2(ip_n-E_{\bf k}^-)^2}=-\sum_{\bf k}\partial_{\xi_{\bf k}}\Big[\frac{\xi_{\bf k}}{2E_{\bf k}}\Big(f(E_{\bf k}^+)-f(E_{\bf k}^-)\Big)\Big]=D.~~~~~
\end{eqnarray}
Further through the integration over the Hartree field, one finally arrives at the effective action of the gap and phase fluctuation:
\begin{equation}
 S_{\rm eff}=\int{dR}\Big\{\sum_{p_n,{\bf k}}\ln[(ip_n-E_{\bf k}^+)(ip_n-E_{\bf k}^-)]-\frac{np_s^2}{2m}-\frac{|\Delta|^2}{U}\Big\}+\int{dt}d{\bf q}\frac{D}{1+2DV_q}\Big(\frac{\partial_{t}\delta\theta_{\bf q}}{2}+\frac{p_s^2}{2m}\Big)^2,
\end{equation}
which gives rise to Eq.~(\ref{action}) by neglecting the high-order mutual interaction between phase fluctuations. 

\subsection{Derivation of thermodynamic equations}
\label{EB}
Through the Euler-Lagrange equation of motion with respect to the SC gap, i.e., $\partial_{|\Delta|}S_{\rm eff}=0$, one has
\begin{eqnarray}
 0=\frac{|\Delta|}{U}+\sum_{p_n,{\bf k}}\frac{|\Delta|}{(ip_n-E^+_{\bf k})(ip_n-E^-_{\bf k})}=\frac{|\Delta|}{U}+\sum_{{\bf k}}\frac{|\Delta|}{2E_{\bf k}}[f(E_{\bf k}^+)-f(E_{\bf k}^-)].
\end{eqnarray}

Through the Euler-Lagrange equation of motion with respect to the long-wavelength phase fluctuation, i.e., $\partial_{\mu}[\frac{\partial{S_{\rm eff}}}{\partial(\partial_{\mu}\delta\theta/2)}]=\partial_{\delta\theta/2}S_{\rm eff}$, in frequency and momentum space, the equation of motion of the NG phase fluctuations reads
\begin{equation}
\Big(\frac{n_sq^2}{2m}-D_q\omega^2\Big)\frac{\delta\theta(\omega,{\bf q})}{2}=0,  \label{PEF}  
\end{equation}
where the superfluid density is determined by 
\begin{eqnarray}
\frac{n_s{\bf p_s}}{m}&=&\frac{n{\bf p}_s}{m}\!+\!{\sum_{p_n{\bf k}}}\frac{2{\bf k}(ip_n\!-\!{\bf k}\!\cdot\!{\bf p}_s/m)}{m(ip_n\!-\!E_{\bf k}^+)(ip_n\!-\!E_{\bf k}^-)}=\frac{{\bf p}_sk_F^2}{m^2}{\sum_{\bf k}}\Big\{-\partial_{\xi_{\bf k}}\Big[\frac{\xi_{\bf k}}{2E_{\bf k}}\Big(f(E_{\bf k}^+)-f(E_{\bf k}^-)\Big)\Big]+\partial_{E_{\bf k}}\Big[\frac{f(E_{\bf k}^+)-f(E_{\bf k}^-)}{2}\Big]\Big\}\nonumber\\
  &=&\frac{{\bf p}_s}{m}\frac{k_F^2}{m}\sum_{\bf k}\frac{|\Delta|^2}{E_{\bf k}}\partial_{E_{\bf k}}\Big[\frac{f(E_{\bf k}^+)-f(E_{\bf k}^-)}{2E_{\bf k}}\Big].
\end{eqnarray}
Equation~(\ref{PEF}) is exactly same as the equation of motion of the NG mode in the literature~\cite{ambegaokar1961electromagnetic,nambu1960quasi,nambu2009nobel,littlewood1981gauge,yang2021theory,sun2020collective}, and  substituting ${\bf p}_s({\bf q})=i{\bf q}\delta\theta({\bf q})/2$ leads to Eq.~(\ref{EOMPS}).

From the equation of motion of the NG phase fluctuation, one can derive the thermal phase fluctuation through the fluctuation dissipation theorem or within Matsubara formalism. The two methods are equivalent, and lead to the exactly same result.   

{\sl Fluctuation dissipation theorem.---}Considering the thermal fluctuation, the dynamics of the phase from Eq.~(\ref{PEF}) is given by
\begin{equation}
\Big(\frac{n_sq^2}{2m}-D_q\omega^2+i\omega{\gamma}\Big)\frac{\delta\theta({\omega,{\bf q}})}{2}={J}_{\rm th}(\omega,{\bf q}).    
\end{equation}  
Here, we have introduced a thermal field ${\bf J}_{\rm th}(\omega,{\bf q})$ that obeys the fluctuation-dissipation theorem~\cite{Landaubook}:
\begin{equation}
\langle{J}_{\rm th}(\omega,{\bf q}){J}^*_{\rm th}(\omega',{\bf q}')\rangle=\frac{(2\pi)^3\gamma\omega\delta(\omega-\omega')\delta({\bf q-q'})}{\tanh(\beta\omega/2)},   
\end{equation}
and $\gamma=0^+$ is a phenomenological damping constant. From this dynamics, the average of the phase fluctuations is given by
\begin{eqnarray}
  \langle{p_{s,\parallel}^2}\rangle&=&\int\frac{d\omega{d\omega'}d{\bf q}d{\bf q'}}{(2\pi)^6}\frac{({\bf q}\cdot{\bf q'})\langle{J}_{\rm th}(\omega,{\bf q}){J}^*_{\rm th}(\omega',{\bf q}')\rangle}{\big[D_q\big(\omega^2\!-\!\omega^2_{\rm NG}({\bf q})\big)\!-\!i\omega\gamma\big]\big[D_{q'}({\omega'}^2\!-\!\omega^2_{\rm NG}({\bf q'})\big)\!+\!i\omega'\gamma\big]}=\int\frac{d\omega{d{\bf q}}}{(2\pi)^3}\frac{q^2\gamma\omega/\tanh(\beta\omega/2)}{\big[D_q\big(\omega^2\!-\!\omega^2_{\rm NG}({\bf q})\big)\big]^2+\omega^2\gamma^2}\nonumber\\
&=&\int\frac{{d{\bf q}}}{(2\pi)^2}\frac{q^2}{2D_q\omega_{\rm NG}(q)}[2n_B(\omega_{\rm NG})+1].\label{EEEE}
\end{eqnarray}

{\sl Matsubara formalism.---}Within the Matsubara formalism, by mapping into the imaginary-time space, from Eq.~(\ref{PEF}),  the thermal phase fluctuation reads~\cite{abrikosov2012methods}
\begin{eqnarray}
  \langle{p_{s,\parallel}^2}\rangle&\!=\!&\int\frac{d{\bf q}}{(2\pi)^2}q^2\Big[\Big\langle\Big|\frac{\delta\theta^*(\tau,{\bf q})}{2}\frac{\delta\theta(\tau,{\bf q})}{2}e^{-\int^{\beta}_{0}d\tau{d{\bf q}}D_q{\delta\theta^*(\tau,{\bf q})}(\omega_{\rm NG}^2-\partial_{\tau}^2){\delta\theta(\tau,{\bf q})}/4}\Big|\Big\rangle\Big]\nonumber\\
  &\!=\!&\int\frac{d{\bf q}}{(2\pi)^2}q^2\Big[\frac{1}{\mathcal{Z}_0}{\int}D\delta\theta{D\delta\theta^*}\frac{\delta\theta^*(\tau,{\bf q})}{2}\frac{\delta\theta(\tau,{\bf q})}{2}e^{-\int^{\beta}_{0}d\tau{d{\bf q}}D_q{\delta\theta^*(\tau,{\bf q})}(\omega_{\rm NG}^2-\partial_{\tau}^2){\delta\theta(\tau,{\bf q})}/4}\Big]\nonumber\\
   &=&\!\!\!\int\frac{d{\bf q}}{(2\pi)^2}q^2\frac{1}{\mathcal{Z}_0}{\int}D\delta\theta{D\delta\theta^*}\delta_{J^*_{\bf q}}\delta_{J_{\bf q}}e^{-\int^{\beta}_{0}d\tau{d{\bf q}}[D_q{\delta\theta^*(\tau,{\bf q})}(\omega_{\rm NG}^2-\partial_{\tau}^2){\delta\theta(\tau,{\bf q})}/4+J_{\bf q}\delta\theta({\bf q})/2+J^*_{\bf q}\delta\theta^*({\bf q})/2]}\Big|_{J=J^*=0}\nonumber\\
  &=&\!\!\!\int\frac{d{\bf q}}{(2\pi)^2}\frac{q^2}{D_q}\delta_{J^*_{\bf q}}\delta_{J_{\bf q}}\exp\Big\{-\int_0^{\beta}{d\tau}\sum_{\bf q'}J_{\bf q'}\frac{1}{\partial_{\tau}^2-\omega_{\rm NG}^2}J^*_{\bf q'}\Big\}\Big|_{J=J^*=0}=-\int\frac{d{\bf q}}{(2\pi)^2}\frac{1}{\beta}\sum_{\omega_n}\frac{q^2}{D_q}\frac{1}{(i\omega_n)^2\!-\!\omega_{\rm NG}^2}\nonumber\\
 &=&\int\frac{{d{\bf q}}}{(2\pi)^2}\frac{q^2}{2D_q\omega_{\rm NG}(q)}[2n_B(\omega_{\rm NG})+1],
\end{eqnarray}
which is exactly the same as the one in Eq.~(\ref{EEEE}) obtained via fluctuation dissipation theorem. Here, $\omega_n=2n\pi{T}$ represents the Bosonic Matsubara frequencies; ${J_{\bf q}}$ denotes the generating functional and $\delta{J_{\bf q}}$ stands for the functional derivative.  

The fundamental derivation of BKT fluctuations and the corresponding RG equations  has been well established in the literature and will not be repeated here.

\subsection{Anderson theorem}
\label{ED}

To facilitate the reading and understanding of the texts and supplementary materials, we reproduce here the well-known the Anderson theorem~\cite{anderson1959theory,suhl1959impurity,skalski1964properties,andersen2020generalized}. We start with the Gorkov equation in Matsubara presentation:
\begin{equation}\label{Dyson}
[ip_n-H_0({\bf k})-\Sigma(ip_n,{\bf k})]G(ip_n,{\bf k})=1,
\end{equation}
where the Bogoliubov–de Gennes Hamiltonian $H_0({\bf k})=\xi_{\bf k}\tau_3+\Delta_0\tau_1$ and the self-energy of the electron-impurity interaction $V_{\bf kk'}$ reads\cite{skalski1964properties}
\begin{equation}
\Sigma(ip_n,{\bf k})=c_i\sum_{{\bf k}'}V_{\bf kk'}\tau_3G(ip_n,{\bf k}')\tau_3V_{\bf k'k}.  
\end{equation}

Without the disorder, the bare Green function of the conventional BCS superconductors is established as
\begin{equation}\label{bare}
G_0(ip_n,{\bf k})=\frac{ip_n+\xi_{\bf k}\tau_3+\Delta_0\tau_1}{(ip_n)^2-\xi_{\bf k}^2-\Delta_0^2}. 
\end{equation}  
With the disorder, to self-consistently calculate the Green function from Eq.~(\ref{Dyson}), one can consider a renormalized equation as~\cite{anderson1959theory,suhl1959impurity,skalski1964properties,andersen2020generalized}
\begin{equation}
ip_n-H_0(k)-\Sigma(ip_n,{\bf k})=i{\tilde p}_n-{\tilde H}_0(k)=i{\tilde p}_n-\xi_{\bf k}\tau_3-{\tilde \Delta}_0\tau_1,\label{SEM_renormalization}  
\end{equation}
which leads to the Green function
\begin{equation}
G(ip_n,k)=\frac{i{\tilde p}_n\tau_0+\xi_{\bf k}\tau_3+{\tilde \Delta_0}\tau_1}{(i{\tilde p}_n)^2-\xi_{\bf k}^2-{\tilde \Delta}_0^2}. \label{SEM_renormalizedGreenfunction}
\end{equation}
Substituting Eq.~(\ref{SEM_renormalizedGreenfunction}) to Eq.~(\ref{SEM_renormalization}), one has 
\begin{eqnarray}
  {\tilde p}_n&=&p_n+\Gamma_0\frac{{\tilde p}_n}{\sqrt{({\tilde p}_n)^2+\Delta_0^2}},\\
  {\tilde \Delta}_0&=&\Delta_0+\Gamma_0\frac{{\tilde \Delta}_0}{\sqrt{({\tilde p}_n)^2+\Delta_0^2}},
\end{eqnarray}
where $\Gamma_0=c_i\pi{N(0)}{\int}d\Omega_{\bf k'}|V_{\bf k_Fk_F'}|^2$. From equations above, it can demonstrated that $p_n/\Delta_0={\tilde p}_n/{\tilde \Delta}_0$.

Consequently, from the gap equation $\Delta_0=-gT\sum_{n,{\bf k}}{\rm Tr}[G(ip_n,{\bf k})\tau_1/2]$, one finds
\begin{equation}
\Delta_0=gT\sum_{n}\int{d{\bf k}}\frac{{\tilde \Delta_0}}{\xi_{\bf k}^2+{\tilde \Delta_0}^2+({\tilde p}_n)^2}=\sum_n\frac{gTN(0)\pi}{\sqrt{1+({\tilde p}_n/{\tilde \Delta}_0)^2}}=\sum_n\frac{gTN(0)\pi}{\sqrt{1+({p}_n/{\Delta}_0)^2}},  
\end{equation}
which is exactly same as the one without the disorder. This indicates that the SC gap equation is insensitive to the non-magnetic impurities, i.e., the SC gap equation experiences a null renormalization by non-magnetic impurities~\cite{anderson1959theory,suhl1959impurity,skalski1964properties,andersen2020generalized}.  

  \begin{figure}[htb]
  {\includegraphics[width=10.6cm]{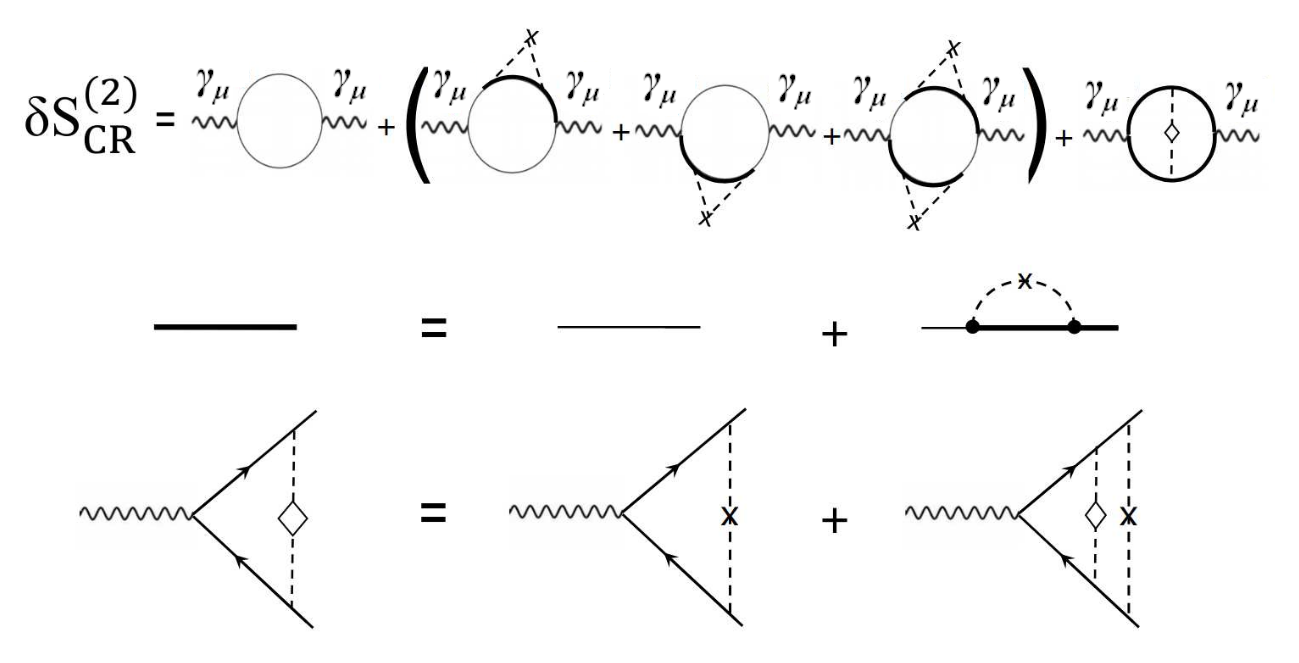}}
  \caption{Diagrammatic formalism for the current–current correlation. The first diagram on the right-hand side of $\delta S_{\rm CR}^{(2)}$ denotes the bare current–current correlation, while the second and third diagrams represent the Born and vertex corrections due to impurity scattering, respectively. Dashed lines with a cross and a diamond denote the impurity potential and the ladder-type series of impurity scatterings. The wavy line corresponds to the external field, and $\gamma_{\mu}$ represents the vertex function, i.e., the four-vector current $\gamma_{\mu}=[\tau_3,\hat{\bf k}/m]$~\cite{schrieffer1964theory,nam1967theory,nambu2009nobel}. Thin solid lines denote the bare SC Green’s functions, while thick solid lines denote the impurity-renormalized SC Green’s functions.}    
\label{Sfigyc}
  \end{figure}

In contrast, the calculation of the superfluid density arises from the current–current correlation function, where vertex corrections due to impurity scattering necessarily appear (Fig.~\ref{Sfigyc}). These corrections vanish in the equilibrium case but become essential for evaluating dynamical or fluctuating responses. Consequently, the superfluid density is renormalized by disorder, in sharp contrast to the robustness of the SC gap.  This can be directly seen from the Mattis–Bardeen formula in the dirty limit, where the penetration depth (or equivalently the superfluid stiffness) depends explicitly on the scattering rate through~\cite{mattis1958theory,tinkham2004introduction} 
\begin{equation}
1/\lambda^2(0)\propto|\Delta_0|ne^2\tau/m,     
\end{equation}
or equivalently 
\begin{equation}
\frac{n_s(T=0)}{n}\propto\frac{l}{\xi}~~~~~~\text{(dirty limit)}. 
\end{equation}
 Clearly, while the gap $|\Delta_0|$ remains insensitive to nonmagnetic impurities, the   superfluid stiffness carries the dependence on the quasiparticle scattering time $\tau$, demonstrating how disorder suppresses the superfluid density without affecting the gap amplitude.  

A fully rigorous treatment would require a microscopic analysis of scattering processes by various theoretical approaches using Gorkov theory~\cite{abrikosov2012methods}, Eilenberger transport~\cite{eilenberger1968transformation,PhysRevLett.25.507,PhysRevB.99.224511}, gauge-invariant kinetic equations~\cite{yang2018gauge}, and diagrammatic formulations incorporating {\sl vertex} corrections to the current–current correlation~\cite{PhysRevB.99.224511,PhysRevB.106.144509}, which we do not repeat here.\\

Importantly, while the SC gap equation itself is not directly renormalized by nonmagnetic impurities, consistent with Anderson’s theorem,  disorder can nevertheless affect the gap equation   \emph{indirectly} by suppressing the phase stiffness and thereby enhancing NG phase fluctuations, whose zero-point and thermal contributions feed back into the gap equation. This mechanism goes beyond the conventional Anderson theorem.

\section{Critical field}

{It has been established that the Nernst signal in 2D superconductors is dominated by vortex physics: a temperature gradient exerts a transverse force on mobile vortices, generating an electric field according to Josephson’s relation. The strength of the Nernst response is therefore controlled by the density and mobility of vortices, which, in turn, are governed by the degree of phase rigidity. As the phase stiffness is reduced by renormalization from NG fluctuations and BKT vortex–antivortex proliferation, the system enters a strongly fluctuating regime with abundant mobile vortices, leading to an enhanced Nernst signal. Conversely, once the superconducting gap collapses, the phase stiffness vanishes and vortices lose their superconducting character, causing the
Nernst signal to disappear.  Thus, the experimentally observed onset of superconductivity (critical density $n_c$), the peak position of the Nernst signal (the density $n_p$ at which the signal maximizes), and the termination field $H_c(0)$ of the Nernst response directly track the fluctuation physics captured by our equilibrium theory.}

{In this part, we derive the critical field $H_{c2}(0)$ here.  Specifically, to cast the phase-gradient energy into the form
  needed to analyze the orbital pair-breaking mechanism, we define $\Psi(\mathbf{R}) \equiv e^{i\delta\theta(\mathbf{R})}$. Then, at zero temperature and in the presence of a vector potential $\mathbf{A}$, 
the energy of the superconducting state can be written, within our phase–only framework, as
\begin{equation}
E = \int d\mathbf{R}\,
\left[
  \frac{\hbar^2 n_s}{2m_e}
  \Big|\Big(\frac{\nabla}{2} - i\frac{e}{\hbar}\mathbf{A}\Big)\Psi(\mathbf{R})\Big|^2
  - D|\Delta(0)|^2\,|\Psi(\mathbf{R})|^2
\right],
\end{equation}
  where the first term represents the phase–gradient energy $\int{d{\bf R}}\frac{n_sp_s^2}{2m_e}=\int{d{\bf R}}\frac{n_s}{2m_e}(\frac{\nabla\delta\theta}{2})^2$ associated with the renormalized superfluid density $n_s$, and the second term is the condensation energy with the density of state $D$. 
Varying this functional with respect to $\Psi^*$ yields the linear equation
 \begin{equation}
-\frac{\hbar^2 n_s}{8m_e}
\left(\nabla - i\frac{2e}{\hbar }\mathbf{A}\right)^2\Psi(\mathbf{R})
-D|\Delta(0)|^2\Psi(\mathbf{R}) = 0.
\end{equation}
For a uniform perpendicular field, $\nabla\times\mathbf{A}=H\hat{\mathbf{z}}$, 
the kinetic operator (i.e., the first term) on the left-hand side has Landau–level eigenvalues
\begin{equation}
\frac{\hbar^2 n_s}{8m_e}\,\frac{n+1/2}{\ell_B^2},
\qquad 
\ell_B^2 = \frac{\Phi_0}{2\pi H},
\end{equation}
with $\Phi_0 = h/2e$ the flux quantum. 
Superconductivity is destroyed when the lowest Landau level ($n=0$) softens, 
which leads to the condition
\begin{equation}
\frac{\hbar^2 n_s}{16m_e}\,\frac{1}{\ell_B^2}
=D|\Delta(0)|^2.
\end{equation}
Using $\ell_B^2 = \Phi_0/(2\pi H_{c2}(0))$, we obtain
\begin{equation}
H_{c2}(0)
= \frac{8D\,m_e\,\Phi_0}{\pi\hbar^2}\,
\frac{|\Delta(0)|^2}{n_s(0)}=\frac{4m_e^2\,\Phi_0}{\pi^2\hbar^4}\,
\frac{|\Delta(0)|^2}{n_s(0)},
\end{equation}
which is the expression used in our analysis of the upper critical field. It should be noted that this derivation follows the standard procedure for obtaining $H_{c2}$, and demonstrates that the resulting dependence 
\begin{equation}
H_{c2}(0)\propto\frac{m_e^2\,\Phi_0}{\hbar^4}\, \frac{|\Delta(0)|^2}{n_s(0)},
\end{equation}
is a universal consequence of the orbital pair breaking effect from external magnetic field. }

\section{Simulation treatment}
\label{secst}

In the specific simulation, a subtle point arises regarding the treatment of the zero-point oscillations of the bosonic NG-mode excitations in Eq.~(\ref{BET}). If not handled properly, this contribution can violate fundamental principles. For instance, in the absence of disorder, the finite expectation value $\langle p_{s,\parallel}^2 \rangle$ generated by the zero-point term $S_{\rm zo}$ in Eq.~(\ref{BET}) would imply a reduced superfluid density $n_s(T=0)<n$ in Eq.~(\ref{SD}), in contradiction with Galilean invariance. As discussed in Refs.~\cite{PhysRevB.64.140506,PhysRevB.69.184510}, this inconsistency for superfluid density can be resolved by introducing a shift of the chemical potential.

In practice, this issue related to the bosonic  excitation is not unique to superconductivity but is also known in the {\sl ab initio} calculations of lattice vibrations within density functional theory. There, the effect of zero-point oscillations is typically absorbed into the renormalized model parameters~\cite{verdi2023quantum,wu2022large}, so that subsequent phonon excitations are described solely by the Bose distribution $n_B(x)$. This approach is consistent with the general field-theoretical treatment of zero-point oscillations~\cite{peskin2018introduction}, where the vacuum is not strictly empty but is continuously populated by zero-point fluctuating excitations. In doing so, the treatment enforces both the Ward identity and the $f$-sum rule, and is essentially equivalent to field-theoretical normal ordering~\cite{peskin2018introduction}, in which the vacuum-related, temperature-independent contribution is absorbed into the ground-state parameters.  Such a procedure has proven effective in the study of lattice dynamics, particularly in quantum paraelectrics~\cite{verdi2023quantum,muller1979srti,li2019terahertz,cheng2023terahertz,wu2022large,74d5-4hsw}.  We thus adopt the same strategy here.  Concretely, we subtract the zero-point contribution $S_{\rm zo}$ in the phase-fluctuation generation of Eq.~(\ref{SD}) by incorporating it into the pairing potential, i.e., by renormalizing $U$ to $\tilde U$. The quantities $\tilde U$ and $\langle{p_{s,\parallel}^2}\rangle=S_{\rm th}$ then replace $U$ and $\langle{p_{s,\parallel}^2}\rangle=S_{\rm th}+S_{\rm zo}$ in the framework, respectively. The renormalized pairing potential is defined as
\begin{equation}\label{S1}
\frac{1}{\tilde U}={F[p_{s,\parallel}^2=0,|\Delta(T=0)|,T=0]},
\end{equation}
whereas $|\Delta(T=0)|$ is self-consistently determined from the original gap equation including the zero-point oscillation,
\begin{equation}\label{S2}
\frac{1}{U}={F[p_{s,\parallel}^2=S_{\rm zo},|\Delta(T=0)|,T=0]}.
\end{equation}
This construction guarantees that the zero-point renormalization of the gap remains unchanged, in consistency with the previous studies~\cite{PhysRevB.97.054510,yang2021theory,PhysRevB.70.214531,PhysRevB.102.060501} and, in particular, that the superfluid density satisfies $n_s(T=0)\equiv n$ in the clean limit,  in consistency with Galilean invariance and previous studies~\cite{PhysRevB.64.140506,PhysRevB.69.184510}. Importantly, this procedure disentangles the zero-point contribution from genuine thermal fluctuations. The former is absorbed into the renormalized pairing interaction at $T=0$, whereas the latter,
\begin{equation}\label{S3}
\frac{1}{\tilde U} = F\left[p_{s,\parallel}^2 = S_{\rm th}\big(T,n_s(T)\big), |\Delta(T)|, T \right],
\end{equation}
exclusively govern the finite-temperature phase-fluctuation physics, as is physically required. It should be emphasized that within this procedure, all physical quantities at $T=0$ and in the $T=0^+$ limit are mathematically continuous and mutually self-consistent.\\

{Thus, for each doping level (Fermi energy) and disorder strength, our simulation proceeds in two conceptually distinct but self-consistent stages.}

\begin{itemize}
\item {{{\bf Zero-temperature initialization}}}.

{{We first determine the zero-temperature reference state by self-consistently solving the zero-temperature gap equation [Eq.~(\ref{S2})], 
\begin{equation}
\frac{1}{U}={F[p_{s,\parallel}^2=S_{\rm zo},|\Delta(T=0)|,T=0]},
\end{equation}
the zero-point Nambu–Goldstone (NG) phase fluctuations [Eq.~(\ref{zo})],
\begin{equation}
S_{\rm zo}=\int\frac{d{\bf q}}{(2\pi)^2}\frac{q^2}{2D_{q}\omega_{\rm NG}(q)},  
\end{equation}
and the disorder-modified superfluid density [Eq.~(\ref{SD2})],
 \begin{equation}
\frac{n_s}{n}=\frac{1}{1+\xi/l}\int{d\xi_k}\int\frac{{d}\theta_{\bf k}}{2\pi}\frac{|\Delta|^2}{2E^3_{\bf k}}[f(E_{\bf k}^-)-f(E_{\bf k}^+)]_{T=0}=\frac{1}{1+\xi/l}\int{d\xi_k}\int\frac{{d}\theta_{\bf k}}{2\pi}\frac{|\Delta|^2}{2E^3_{\bf k}}\Theta(-E_{\bf k}^-)\Theta(E_{\bf k}^+).   
\end{equation}
These quantities are updated iteratively until convergence is reached, yielding the zero-point–renormalized superconducting gap $|\Delta(T=0)|$. The resulting renormalized gap is then used to fix the effective pairing potential through Eq.~(\ref{S1}), 
\begin{equation}
\frac{1}{\tilde U}={F[p_{s,\parallel}^2=0,|\Delta(T=0)|,T=0]},
\end{equation}
which serves as the input for the finite-temperature calculations.}}

\item {{{\bf Finite-temperature self-consistent loop}}}.

{{Using the pairing potential determined in zero-temperature initialization, we subsequently solve, at each temperature, the finite-temperature gap equation [Eq.~(\ref{S3})],
\begin{equation}
\frac{1}{\tilde U} = F\left[p_{s,\parallel}^2 = S_{\rm th}\big(T,n_s(T)\big), |\Delta(T)|, T \right],
\end{equation}
the thermal NG phase fluctuations [Eq.~(\ref{th})],
\begin{equation}
S_{\rm th}(T)=\int\frac{d{\bf q}}{(2\pi)^2}\frac{q^2n_B(\omega_{\rm NG})}{D_{q}\omega_{\rm NG}(q)}, 
\end{equation}
and the disorder-modified superfluid density [Eq.~(\ref{SD2})] 
 \begin{equation}
\frac{n_s}{n}=\frac{1}{1+\xi/l}\int{d\xi_k}\int\frac{{d}\theta_{\bf k}}{2\pi}\frac{|\Delta|^2}{2E^3_{\bf k}}[f(E_{\bf k}^-)-f(E_{\bf k}^+)].   
\end{equation}
in the same iterative manner. This procedure yields the 
finite-temperature gap $|\Delta(T)|$, and in particular, bare finite-temperature superfluid density $n_s(T)$, which is then used as the input to the BKT renormalization-group equations [Eqs.~(\ref{BKT1})–(\ref{BKT2})],
\begin{equation}
\frac{dK}{dl}=-K^2g^2~~~\text{and}~~~\frac{dg}{dl}=(2-K)g,
\end{equation}
and
\begin{equation}
K(l=0)=\frac{\pi}{k_BT}\frac{\hbar^2{n_s}}{4m}~~~\text{and}~~~g(l=0)=2\pi{e}^{-\mu_v(T)/(k_BT)}
\end{equation} 
Integrating the BKT flow to $l\rightarrow\infty$ yields a renormalized superfluid density 
\begin{equation}
{\bar n}_s=\frac{4m_ek_BT}{\pi\hbar^2}K(l=\infty).
\end{equation}}}

\end{itemize}

{The momentum cutoff $q_{\rm cutoff}$ in the bosonic integrals of the NG-mode excitation is chosen as $1/\xi$, in consistency with the previous studies~\cite{yang2021theory,PhysRevB.97.054510,PhysRevB.70.214531}, and this is because that the quasiparticle scattering processes effectively limit the phase coherence at length scales shorter than the SC coherence length. As a result, NG phase fluctuations with wavelengths shorter than the SC coherence length are strongly suppressed and do not contribute appreciably to the thermodynamic dynamics.}

{{From the resulting finite-temperature gap $|\Delta(T)|$ and the renormalized superfluid density $\bar n_s(T)$, we determine the full set of superconducting critical properties. In particular, the superconducting transition temperature $T_c$ is determined from the condition ${\bar n}_s(T_c)=0$ while the gap‐closing temperature $T_{\rm os}$ is obtained by $|\Delta(T=T_{\rm os})|=0$.}} \\

{\bf For WTe$_2$}: The electron density $n_e=2n$ (reflecting the twofold valley degeneracy at $\pm Q$) and the hole density $n_h$ are determined by the Fermi energy $E_F$ and the negative band gap $E_g$ within the parabolic-band 2DEG approximation: $n=2DE_F$ with the electron density of states $D=m_e/(2\pi\hbar^2)$ and $n_h=2m_h/(2\pi\hbar^2)\times(|E_g|-E_F)\theta(|E_g|-E_F)$. The energy zero is chosen at the conduction-band minimum; in the pristine case, the Fermi level thus coincides with the conduction-band edge, consistent with experimental observations showing that monolayer WTe$_2$ is effectively $p$-type at zero gate bias.

Moreover, as discussed in the main text, to approximately model the competition with the excitonic insulator phase (which in monolayer WTe$_2$ is sufficiently strong to dominate over the electron SC channel~\cite{Jia2022,fatemi2018electrically,sajadi2018gate,song2024unconventional,PhysRevResearch.7.013224}), we assume that each hole binds with one conduction electron to form an exciton, thereby depleting the electrons available for SC pairing.  This procedure is implemented specifically by reducing superconducting-pairing-active electronic states, namely the effective density of states entering the electron–SC gap equation and the superfluid density by a factor of $(n_e-n_h)/n_e$ in the self-consistent numerical simulation. {This treatment is a commonly used technique when modeling competing 
orders in strongly correlated superconducting systems (e.g., cuprates), which is an effective way to capture the dominant impact of a many-body competition from other 
ordered phase on superconducting pairing when extended coexistence is not expected. It should be emphasized that a comprehensive calculation of excitonic insulating order is not the focus of the present work; this excitonic instability is included as a competing ingredient to capture the experimentally observed suppression of superconductivity at low doping, rather than as a primary subject of superconductivity calculation. Consequently, the disappearance of superconductivity at the critical doping level is because the electronic states
participating in SC pairing are effectively exhausted, as the system crosses into an excitonic
insulating state that does not coexist with superconductivity. This physical picture is consistent with many-body theoretical analyses~\cite{PhysRevB.71.224502} showing that
excitonic order and superconductivity generally do not coexist over an extended parameter regime.  Instead, the two orders are largely mutually exclusive, with at most a
very narrow coexistence region, particularly in systems where the electron–hole pairing interaction 
is stronger than the superconducting pairing interaction. Notably, it is
also supported by experimental observations, for example as shown in Ref.~\cite{Tang2023AmbipolarMoTe2}, where the suppression of superconductivity coincides with the onset of an excitonic insulating
phase.}

\begin{center}
\begin{table}[htb]
\caption{Parameters used in simulation. The electron and hole effective masses in monolayer WTe$_2$ are obtained from our DFT calculations (Sec.~\ref{DFT}), $m_e\approx0.33m_0$ and $m_h\approx0.53m_0$. The pairing potential and the energy range of the attractive interaction  are not available in the literature. Given that the experimentally observed SC transition temperatures largely lie below 1~K, we restrict our analysis to the vicinity of the Fermi surface within an energy cutoff of $E_c \sim 2$~K. The pairing potential $U$ relevant to the experiments in Refs.~\cite{fatemi2018electrically,sajadi2018gate,song2024unconventional,PhysRevResearch.7.013224} is determined by fitting the SC transition temperature in the high-density regime (mean-field regime).}
\label{divi} 
\begin{tabular}{l l l l l l}
    \hline
    \hline
    Simulation parameters used in the main text &\quad~Fig.~1&\quad~Figs.~2(b), inset and~Fig.~3&\quad~Fig.~2(c) &\quad~inset of Fig.~2(c)&\quad~Fig.~2(d)   \\
    \hline \\
    ${DU}/{2}$&\quad~$0.462$&\quad\quad\quad~$0.554$&\quad~$0.40$&\quad\quad\quad~$0.423$&\quad~0.506\\~\\
    $|E_g|$~(meV)&\quad~0&\quad\quad\quad~39&\quad~26&\quad\quad\quad~29&\quad~18\\~\\   
    $\gamma$~(meV)&\quad~0-4&\quad\quad\quad~$3.5$&\quad~$0.8$&\quad\quad\quad~$0.54$&\quad~$0.1$\\~\\  
    Corresponding experiments&\quad~--&\quad\quad\quad~Ref.~\cite{song2024unconventional}&\quad~Ref.~\cite{fatemi2018electrically}&\quad\quad\quad~Ref.~\cite{sajadi2018gate}&\quad~Ref.~\cite{PhysRevResearch.7.013224}
    
    \\~\\
    \hline
    \hline
\end{tabular}\\
\end{table}
\end{center}

{\bf Justification for the sample dependence}

{It is well documented in gate-induced two-dimensional materials that atomically thin systems are highly susceptible to strain, dielectric environment, substrate screening, gating geometry, and fabrication conditions, all of which can significantly renormalize the band structure and phonon-mediated pairing interaction. Consequently, both band gap $E_g$ and the effective attractive interaction $U$ are expected to vary across samples even within nominally the same material platform.}

{The disorder in 2D  materials is exceptionally sensitive to sample preparation, encapsulation, gating history, and environmental exposure. As a result, large variations in the effective  scattering time are routinely observed and are widely recognized in transport measurements on gated monolayer systems. Such sample-to-sample variability is a well-known and intrinsic feature of 2D materials, rather than an indication of unphysical parameter tuning. In addition,  it should also be emphasized that the relevant timescale in the present context is the SC phase-coherence dephasing time that controls phase stiffness and phase fluctuations, rather than the momentum-relaxation time extracted from transport measurements.  It is also not expected to coincide quantitatively with the Drude momentum-relaxation time. Consequently, a resistivity-based estimate does not provide a controlled measure of the timescale relevant to the suppression of superconductivity.  The dephasing time is typically accessed through time- or frequency-resolved measurements, such as microwave response, ac conductivity, or phase-sensitive probes.}

{These sample-dependent variations reflect intrinsic differences in material quality, device fabrication, and experimental conditions, which must ultimately be characterized and constrained by experiment and cannot be eliminated by theory alone. We also note that the parameter variations employed in our work are neither excessive nor unphysical; the pairing interaction $DU/2 \in [0.4,0.55]$, the disorder-induced dephasing rate $\gamma\in[0.1,3.5]~$meV, and the band gap $E_g\in[20,40]~$meV.  Variations of this magnitude (at the meV energy scale for the scattering rate and at the level of tens of meV for the band gap) are commonly observed due to differences in dielectric environment, strain, gating conditions, and sample quality.}

\section{Density-functional-theory calculation}
\label{DFT}

The first-principles density functional theory simulations are performed with projector augmented-wave method implemented in the Vienna ab initio Simulation Package (VASP)~\cite{VASP1,VASP2}. 
The exchange-correlation functional is the Perdew–Burke–Ernzerhof (PBE) type generalized gradient approximation (GGA)~\cite{PBE}. 
The plane wave cutoff energy is set as 600 eV. 
For the full atomic relaxation with a fixed vacuum layer larger than 20~{\AA} and free in-plane lattice parameters, we included the DFT-D3 Van der Waals correction with Becke-Johnson damping~\cite{DFTD3}, following the same settings as in our previous work~\cite{Xingen}. 
The relaxation is converged after all the atomic forces are below 1 meV/\AA, and electronic energy convergence criterion was set to $1\times10^{-7}$ eV.
We used a $\Gamma$-centered k-point grid of $9\times5\times1$ for relaxation and $20\times10\times1$ for static electronic calculations.
We considered 14 valence electrons for the W element and 6 for the Te element.
The relaxed in-plane lattice parameters are 3.453 and 6.262 \AA, respectively.
Spin-orbit coupling (SOC) is considered for static and band structure calculations to account for the band inversion. 

Electron and hole effective masses are extracted from parabolic fitting of the band dispersion $E(k)$ around the conduction band minimum and valence band maximum, respectively.
Both the x- and y-direction effective masses were evaluated, from additional k-lines sampled crossing the corresponding extreme points along orthogonal directions, as shown in Table~\ref{tab:massfit}.  These results indicate nearly isotropic electron effective masses, and lightly anisotropic hole effective masses. 

\begin{table}
\caption{Effective masses of electrons and holes obtained from quadratic fits of band dispersions near the conduction band minimum (CBM) and valence band maximum (VBM) along the x- and y-directions. The quality of the fits is characterized by the coefficient of determination $R^2$, with $R^2 \approx 1$ indicating high reliability.}
\centering
\begin{tabular}{cccccccccccccc}
\hline\hline
\multicolumn{2}{c}{Range}                    & \multicolumn{3}{c}{CBM-x}          & \multicolumn{3}{c}{CBM-y}     & \multicolumn{3}{c}{VBM-x}          & \multicolumn{3}{c}{VBM-y}     \\
\hline
k-range ($\AA^{-1}$) & k-points & m*/m$_0$    & $\Delta E$ (eV)  & R$^{2}$ & m*/m$_0$    & $\Delta E$ (eV)  & R$^{2}$ & m*/m$_0$    & $\Delta E$ (eV)  & R$^{2}$ & m*/m$_0$    & $\Delta E$ (eV)  & R$^{2}$\\
{[}-0.020,+0.020{]} & 19 & 0.324  &  0.0043  &  0.965  &  0.336  &  0.0042  &  0.959  &  0.549  &  0.0025  &  0.960  &  0.516  &  0.0027  &  0.960 \\
{[}-0.030,+0.030{]} & 29 & 0.321  &  0.0107  &  0.971  &  0.333  &  0.0098  &  0.982  &  0.552  &  0.0059  &  0.983  &  0.517  &  0.0063  &  0.982 \\
{[}-0.040,+0.040{]} & 39 & 0.314  &  0.0206  &  0.965  &  0.329  &  0.0181  &  0.990  &  0.557  &  0.0105  &  0.990  &  0.518  &  0.0114  &  0.990 \\
{[}-0.050,+0.050{]} & 49 & 0.305  &  0.0355  &  0.949  &  0.325  &  0.0290  &  0.993  &  0.564  &  0.0163  &  0.994  &  0.520  &  0.0179  &  0.994 \\
\hline\hline
\end{tabular}
\label{tab:massfit}
\end{table}

\section{Potential of hole-SC state}

\begin{figure}[htb]
  {\includegraphics[width=15.7cm]{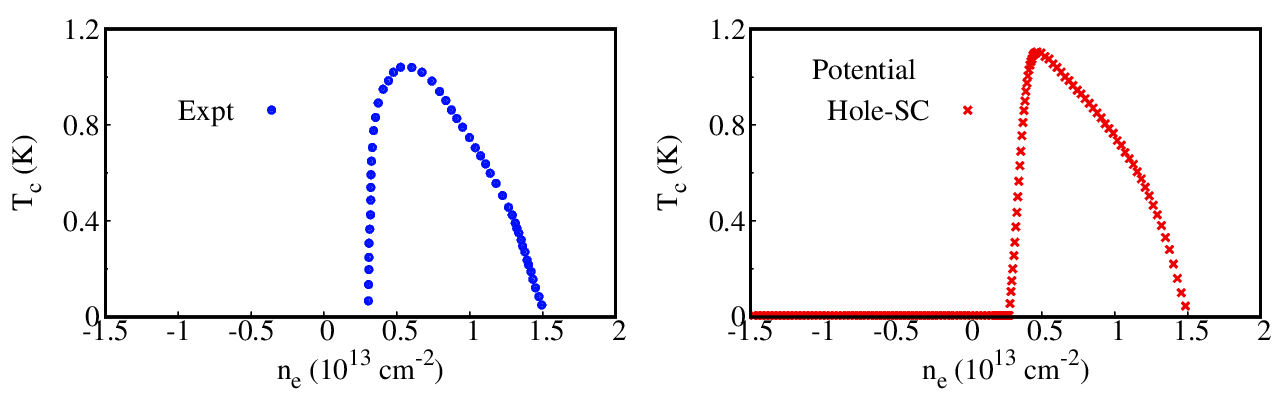}}
  \caption{Possibility of hole superconductivity on the electron-doping side. In the figure, the density $n_e=4DE_F$ with the electron density of states $D=m_e/(2\pi\hbar^2)$. Blue circles in left panel are extracted from the experimentally observed $T_c$ of the  superconductivity that survives at $H=50~$mT~\cite{PhysRevResearch.7.013224}. Red crosses in right panel show the calculated $T_c$ from our toy model. In the simulation, $D_hU_h/2=0.703$ with $U_h$ being the pairing potential of hole-SC state. $\gamma=0.36~$meV,  $E_g=55~$meV and $n_h^*=0.66\times{10^{13}}/\text{cm}^2$. }
\label{figyc4}
  \end{figure}

In this part, we construct a toy model to examine the possibility of the hole-SC state on the electron-doped side.  Specifically, the total hole density is given by
\begin{equation}
n_h = 2D_h(E_g - E_F)\theta(E_g - E_F),
\end{equation}
where $D_h=m_h/(2\pi\hbar^2)$ is the density of states of the hole band, $\theta(x)$ is the step function, and $E_g$ is the  band gap at which holes begin to emerge.  We assume that excitonic pairing sets in only once the hole density exceeds a certain threshold $n_h^*$, beyond which the excitonic phase rapidly preempts the hole-SC state. In the strong-disorder regime, the SC gap of the holes is directly suppressed due to their small phase stiffness, as discussed in the main text, so that $n_h^*=0$, i.e., any finite hole density immediately participates in excitonic pairing. Under weak disorder, however, we can assume $n_h^*>0$, and beyond this point, the hole-sector SC state is rapidly quenched as the excitonic phase preempts it.  In this case, within the SC framework in the main text, by incorporating the effective hole mass ($m_h\approx0.53m_0$) obtained from our DFT calculations and tuning $E_g$, $n_h^*$, the SC pairing potential of the holes and the corresponding disorder strength (weak disorder), we can reproduce the dome-like evolution of the additional SC channel, which closely resembles the experimental observations in a quantitative manner, as shown in Fig.~\ref{figyc4}.

We emphasize, however, that this model is merely an illustrative construct intended solely to visualize a plausible scenario of the hole-SC channel on the electron-doping side, and is included only for heuristic completeness. It offers no conceptual or meaningful advance, as the hole-SC channel depicted here could also be replaced by other unconventional pairing channel without loss of generality. A fully rigorous microscopic analysis would require additional experimental constraints, such as scanning tunneling microscopy (STM) measurements of the SC gap(s) or independent evidence for multiband superconductivity, as well as {\sl ab} initio estimates of the pairing interactions in both the conduction and valence bands. These inputs are essential to clarify the competition, possible coexistence, and mutual coupling among the three candidate many-body states on the electron-doping side: electron-SC, hole-SC, and excitonic-insulator phases, but beyond the scope of the present study.
